\documentclass[useAMs,a4paper]{mn2e}
\usepackage{savesym}
\usepackage{graphicx}
\expandafter\let\csname equation*\endcsname\relax
  \expandafter\let\csname endequation*\endcsname\relax 
\usepackage{subfig}
\usepackage{amsmath}
\usepackage{amssymb}
\usepackage{verbatim}
\usepackage[yyyymmdd,hhmmss]{datetime}
\usepackage{array}
\usepackage{times}
\usepackage[total={17.8cm,24.0cm},centering]{geometry} 
\usepackage{color}

\newcommand{\be}{\begin{equation}}
\newcommand{\beq}{\begin{equation}}
\newcommand{\ee}{\end{equation}}
\newcommand{\eeq}{\end{equation}}
\newcommand{\eea}{\end{eqnarray}}
\newcommand{\bea}{\begin{eqnarray}}

\newcommand{\dd}{\partial}

\newcommand\W {{W^r_{\ \phi}}}
\newcommand\T {{\mathcal{T}}}

\newcommand\Asa{{ASASSN-14li}}
\newcommand\oo {{\rm o}}

\title[Spectral evolution of disc dominated TDEs ]{The spectral evolution of disc dominated tidal disruption events}
\author [Andrew Mummery, Steven A. Balbus]{Andrew Mummery\thanks{E-mail:
andrew.mummery@physics.ox.ac.uk}, {Steven A. Balbus}
\\
Oxford Astrophysics, Denys Wilkinson Building, Keble Road, Oxford, OX1 3RH, United Kingdom}
\begin{document}

\date{}

\pagerange{\pageref{firstpage}--\pageref{lastpage}} \pubyear{2020}

\maketitle

\label{firstpage}

\begin{abstract} 
We perform a detailed numerical and analytical study of the properties of observed  light curves from relativistic thin discs, focussing on observational bands  most appropriate for comparison with tidal disruption events (TDEs).  We make use of asymptotic expansion techniques applied to the spectral emission integral, using time dependent disc temperature profiles {appropriate for solutions of the relativistic thin disc equation}. Rather than a power law associated with {\it bolometric} disc emission $L \sim t^{-n}$, the observed X-ray flux from disc-dominated TDEs will typically have the form of a power law multiplied by an exponential (see eq.\ \ref{helum}).  
While precise details are somewhat dependent on the nature of the ISCO stress and disc-observer orientational angle, the general form of the time-dependent flux is robust and insensitive to the exact disc temperature profile.   We present numerical fits to the UV and X-ray light curves of ASASSN-14li, a particularly well observed TDE.   This  modelling incorporates strong gravity optics.  The full 900 days of \Asa\ X-ray observations are very well fit by a simple relativistic disc model, significantly improving upon previous work.   The same underlying model also fits the final 1000 days of \Asa\ observations in three different  UV bandpasses.    Finally, we demonstrate that the analytic formulae reproduce the properties of full numerical modelling at both UV and X-ray wavelengths with great fidelity. 
\end{abstract}

\noindent

\section{Introduction}

The evolution of bright flares following the tidal destruction of a star passing near the horizon of a supermassive black hole, a so-called tidal disruption event (TDE), comprise a particularly interesting application of relativistic disc theory.     In a recent series of papers (Mummery \& Balbus 2019a,b; hereafter MB1, MB2), the authors have studied the solutions of the  thin disc equation valid in Kerr spacetime geometries.    The models studied were necessarily idealised in many ways: in addition to the standard thin disc assumptions, the models did not include fallback onto the disc or outflows in the form of a jet or wind, the disc was assumed to be confined to the Kerr geometry symmetry plane, and the stress tensor was parameterised as the product of power laws in surface density and radius, in accord with standard $\alpha$-disc modelling.    By stipulating a mathematically stable regime for the stress, these studies did not address the possible presence of Lightman-Eardley viscous instabilities.    Nevertheless, by stripping the problem down to its simplest form an important result emerged:  the choice of disc inner boundary conditions leads to very distinct predictions for the behaviour of the late time ($t$) bolometric luminosity.   

The presence of an innermost stable circular orbit (ISCO) within a relativistic disc implies that the disc must be in effect cut-off, since gas interior to this radius, facing no angular momentum barrier, is able to plunge toward the centre.   A purely viscous stress would arguably vanish at this location (e.g.\ Page \& Thorne 1974), as it does at a free fluid boundary.   Astrophysical discs are much more likely to be dominated by a magnetic stress.   As this requires only correlated (radial and azimuthal) magnetic fields to be present, there is no reason for this quantity to vanish at the ISCO.    This is of astrophysical importance, since the presence or absence of an ISCO stress changes the late time behaviour of the emergent bolometric luminosity, $L(t)$.   Generally, $L(t)$ behaves as a power law $\sim t^{-n}$ at late times.  For a vanishing ISCO stress, $L$ falls off more steeply than $t^{-1}$ (Cannizzo {\it et al}. 1990), whereas for a finite ISCO stress, the fall off is less steep than $t^{-1}$ (Balbus \& Mummery 2018, BM18).    Recent high quality data for  four ``confirmed'' late time X-ray TDEs  (Auchettl, Guillochon, \& Ramirez-Ruiz 2017) all show the latter, more shallow, decline for their light curves.     This is very suggestive, if indirect, evidence for the presence of a finite ISCO stress, which has historically been a somewhat contentious theoretical issue ({Gammie 1999, Krolik 1999, Agol \& Krolik 2000}).  
 
The theoretical luminosity calculations described above are bolometric, whereas observations are always taken over a well-defined instrumental bandpass.    It is of interest therefore to investigate the time dependent emission  from the simple 1D relativistic disc models of MB1 and MB2 over prescribed wavelength intervals.  In this work we both use and generalise asymptotic expansion techniques applied to the disc spectral  integral (Balbus 2014) to derive rather simple, but fully time-dependent, analytic expressions appropriate to UV and X-ray bandpasses. These are very general results with wide applicability. In principle, such calculations, when fitted to observations, can reveal important properties of both the disc and the host black hole.   

As an application of this asymptotic analysis, we present direct data fits to the confirmed TDE \Asa, a particularly well observed TDE first discovered on 22 November 2014 (Holoein {\it et al.} 2016).  The host galaxy of \Asa\ has been identified, and at a distance of 90 Mpc \Asa\ is the closest observed TDE in more than a decade. There are high quality UV (Brown {\it et al.} 2017, Van Velzen {\it et al.} 2019) and X-ray (Bright {\it et al.} 2018) observations of \Asa\, spanning 1300 and 900 days respectively. {Spectrally \Asa\ belongs to a so-called `super-soft' class of TDEs, which are characterised by X-ray spectra that have been modelled by a blackbody profile (Miller {\it et al.} 2015, Brown {\it et al}. 2017). The well-sampled light curves and likely thermal origin of the emitted flux makes \Asa\ an ideal candidate for our analysis. }   In this paper we apply a simple relativistic disc model to this system, fitted to the evolving UV and X-ray light curves. We find that this model not only provides a significantly improved fit to the observed X-ray light curve of ASASSN-14li when compared to other models in the literature, it simultaneously fits the late time ($t > 200$ days) UV emission.   Both the thin disc dynamics and the ray tracing calculations are fully relativistic.

The layout of the paper is as follows. In \S2, we describe the formal problem to be solved, which is to calculate the evolving spectrum from a relativistic thin disc.  In \S3,  we present an asymptotic analysis of the disc spectral integral in a number of physically meaningful limits. The direct fits to the \Asa\ light curves are presented in \S 4, 5 and 6. In \S4,  we describe the underlying relativistic disc model and present fiducial fitted light curves. In \S 5, we examine the effect of varying black hole spin on the best fit light curves. In \S 6 we compare the best fit X-ray light curves of discs evolving with a finite and vanishing ISCO stress. Finally, in \S7 we perform a quantitive comparison of the asymptotic analysis, the best fit numerical light curves and the observed \Asa\ light curves, before concluding in \S8.

\section{Relativistic disc spectrum }

\subsection{Dynamical equation}
The underlying disc model describes the evolution of the azimuthally-averaged, height-integrated disc surface density $\Sigma (r, t)$.    Standard cylindrical Kerr geometry Boyer-Lindquist coordinates are used: $r$ (radius), $\phi$ (azimuth), $z$ (height), $t$ (time), and $\text{d}\tau$ (invariant line element).  The contravariant four velocity of the disc fluid is denoted $U^\mu$ (related to coordinate $x^\mu$ by $U^\mu=\text{d}x^\mu/\text{d}\tau$); its covariant counterpart is $U_\mu$.  The specific angular momentum corresponds to $U_\phi$, a covariant quantity.     We assume that there is an anomalous stress tensor present, $\W$, due to low-level disk turbulence.   The stress is a measure of the correlation between the fluctuations in $U^r$ and $U_\phi$ (Balbus 2017), and could also include correlated magnetic fields.  As its notation suggests, $\W$ is a mixed tensor of rank two.    

It is convenient to introduce the quantity $\zeta$,
\beq\label{z}
\zeta \equiv \sqrt{g}\Sigma \W / U^0=  r \Sigma \W/U^0,
\eeq
where $g>0$ is the absolute value of the determinant of the (mid-plane) Kerr metric tensor $g_{\mu\nu}$.  The Kerr metric  describes the spacetime external to a black hole of mass $M$ and angular momentum $J$.  For our choice of (midplane) coordinates, $\sqrt{g}=r$.   The ISCO radius, inside of which the disc is rotationally unstable, is denoted as $r_I$.  Other notation is standard: the gravitation radius is  $r_g = GM/c^2$, and the black hole spin parameter is $a = J/Mc$. 

Under these assumptions, the governing equation for the evolution of the disc may generally be written (Eardley \& Lightmann 1974; Balbus 2017):
\beq\label{fund}
{\dd \zeta\over \dd t} =  \mathcal{W} {\dd\ \over \dd r}\left({U^0\over U'_\phi}    {\dd \zeta \over \dd r} \right), 
\eeq
where the primed notation denotes a radial gradient, and we have defined the stress-like quantity
\beq
\mathcal{W} \equiv  {1 \over (U^0)^2} \left(\W + \Sigma {\dd\W\over \dd \Sigma} \right)  .
\eeq
The properties of this equation and its time-dependent solutions have been discussed in some detail in BM18, MB1, and MB2.  

\subsection{Spectral integral}
\begin{figure}
  \includegraphics[width=.5\textwidth]{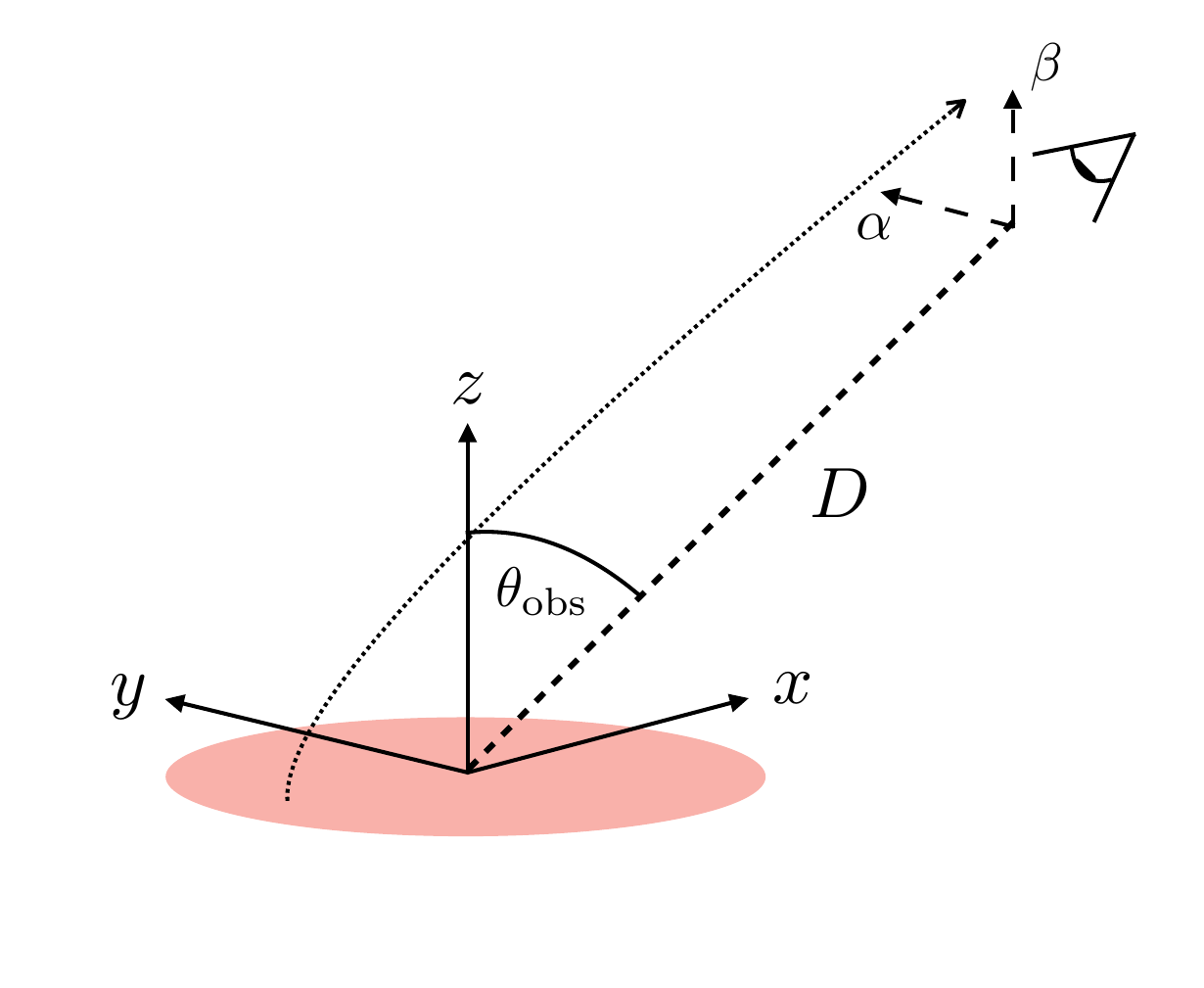} 
 \caption{Ray tracing geometry.  The coordinates $\alpha$ and $\beta$ lie in the observer plane; $x$ and $y$ lie in the disc plane.   A schematic photon trajectory from the inner disc is shown.  The observer-disc inclination angle is denoted $\theta_{\rm obs}$.} 
 \label{fig1}
\end{figure}

The dominant $r\phi$ component of the turbulent stress tensor $\W$ serves to transport angular momentum outward as well as to extract the free energy of the disc shear, which is then thermalised and radiated from the disc surface.   In standard $\alpha$-disc modelling, which we follow here, both the the extraction and the dissipation are assumed to be local processes.
With these assumptions, the profile of the disc surface temperature $T$ is given by (Balbus 2017)
\beq
\sigma T^4 =  - \frac{U^0U^\phi}{2r}  ~ (\ln\Omega) '~ \zeta(r,t)  ,
\label{temperature}
\eeq
where $\sigma$ is the  Stefan-Boltzmann constant and 
\beq
\Omega \equiv {\text{d}\phi\over \text{d}t} = {U^\phi\over U^0}.
\eeq
The specific flux density $F_\nu$ of the disc radiation, as observed by a distant observer at rest (subscript ${\rm o}$), is given by
\beq
F_{\nu}(\nu_{\rm o}) = \int I_\nu (\nu_\oo) \, \text{d}\Theta_{{\oo}} .
\eeq 
Here, $\nu_\oo$ is the photon frequency and $I_\nu(\nu_\oo)$ the specific intensity,  both measured at the location of the distant observer O.   The differential element of solid angle subtended on the observer's sky by the disk element is $\text{d}\Theta_{{\oo}}$. 
Since $I_\nu / \nu ^3$ is a relativistic invariant (e.g. Misner, Thorne, \& Wheeler 1973), we may write
\beq
F_{\nu}(\nu_\oo) = \int f^3 I_\nu (\nu_e) \, \text{d}\Theta_{{\oo}},
\eeq 
where we define the frequency ratio factor $f$ as the ratio of $\nu_\oo$ to the emitted local rest frame frequency $\nu_e$:
\begin{equation}\label{redshift}
f(r,\phi) \equiv \frac{\nu_{\oo}}{\nu_e} = {p_\mu U^\mu\ ({\rm Ob})\over p_\lambda U^\lambda\ ({\rm Em})}=\frac{1}{U^{0}} \left[ 1+ \frac{p_{\phi}}{p_0} \Omega \right]^{-1} ,
\end{equation}
where (Ob) and (Em) refer to observer and emitter, respectively.   The covariant quantities $p_\phi$ and $-p_0$ (on the far right) correspond to the angular momentum and energy of the {\em emitted photon} in the local rest frame.   These may be conveniently regarded as constants of the motion for a photon propagating through the Kerr metric.    Except for special viewing geometries, these quantities must in general be found by numerical ray tracing calculations (see  Appendix \ref{appendixA}). 

The disc is assumed to be a multi-temperature black body, with surface temperature $T=T(r,t)$ given by equation (\ref{temperature}).  The specific intensity of the locally emitted radiation is then given by the Planck function $B_\nu$
\beq\label{planck}
I_\nu(\nu_e) = B_\nu(\nu_e,T) = \frac{2h\nu_e^3}{c^2} \left[ \exp\left( \frac{h\nu_e}{k_B T} \right) - 1\right]^{-1} .
\eeq
For an observer at a large distance $D$ from the source, the differential solid angle into which the radiation is emitted is
\beq
 \text{d}\Theta_{{\oo}} = \frac{\text{d}\alpha \, \text{d} \beta}{D^2} ,
\eeq 
where $\alpha$ and $\beta$ are the impact parameters at infinity (Li \textit{et al}. 2005).  See {Fig.\  (\ref{fig1})}  for further details.   
The observed flux from the disc surface ${\cal S}$ is therefore formally given by
\beq\label{flux}
F_\nu(\nu_\oo,t) =  \int_{\cal S} {f^3 B_\nu (\nu_\oo/f , T)}\,  \text{d}\Theta_\oo . 
\eeq
To compare our results with \textit{Swift} X-ray observations, we require $F_X$, the total flux observed between the telescope's bandpass at 0.3 and 10 keV.    This is easily calculated by integrating (\ref{flux}) over the corresponding (observer) frequency range:
\beq\label{FX}
F_X(t) =  \int_{\nu_l}^{\nu_u}\int_{\cal S} {f^3 B_\nu (\nu_\oo/f , T)} ~ \text{d}\Theta_\oo
\ \text{d}\nu_\oo .
\eeq 
{Where $\nu_l$ and $\nu_u$ correspond to the lower (0.3 keV) and upper (10 keV) frequencies of the {\it Swift} telescope X-ray band}. 
Note that $f$ will generally depend upon $\alpha$ and $\beta$.   With $T$ given by equations (\ref{fund}) and (\ref{temperature}), and ray tracing calculations determining  $f(\alpha,\beta)$, the observed spectrum may be obtained with relativistic effects (kinematic and gravitational Doppler shifts, and gravitational  lensing) included. These are needed for detailed comparison with observations at X-ray wavelengths. {Finally, we note that radiative transfer physics such as absorption and Compton scattering of photons as they pass through the disc atmosphere has been ignored, a simplification that is often made.   For spectrally `super-soft'  TDEs like \Asa\ these corrections are likely to be modest, as any large effects would result in noticeable deviations from simple integrated blackbody spectra.   A more detailed radiative treatment is of interest, but will be left for a later investigation.}

\section{Asymptotic analysis }\label{analytic}

\subsection{Effective temperature}
The canonical thermal accretion disc spectrum is by now well known:  a low frequency Rayleigh-Jeans tail, an intermediate frequency power law section (arising from a superposition of blackbody emission peaks), and a high frequency,  quasi-Wien tail.    The relativistic case considered here is broadly similar, but with important distinctions in detail.  

Begin by recasting the equations (\ref{flux}) and  (\ref{FX}) into a somewhat more classical form by defining an ``effective temperature'' $\widetilde T$ by
\beq\label{Teff}
\widetilde T(\alpha, \beta, t) = f(\alpha, \beta) ~T\big( r(\alpha,\beta), t \big) .
\eeq
This leads to 
\beq\label{flux2}
F_\nu(\nu_\oo,t) = 
{\frac{2h\nu_o^3} {c^2} \int_{\cal S}  \left[ \exp\left( {h\nu_\oo}/{k_B \widetilde T} \right) - 1\right]^{-1} } ~ \text{d}\Theta_\oo , 
\eeq
and 
\beq\label{FX2}
F_X(t) =   \int_{\nu_l}^{\nu_u}\int_{\cal S} \ \ 
\frac{2h\nu_\oo^3}{c^2}  \frac {\text{d}\Theta_\oo\,  \text{d}\nu_\oo}  {\exp( {h\nu_\oo}/{k_B \widetilde T} ) - 1 }.
\eeq
The high photon energy behaviour $(h\nu_\oo \gg k_B \widetilde T)$ of equation 
(\ref{flux2}) may be found analytically by an asymptotic expansion of these integrals, without any detailed knowledge of the underlying disc model (Balbus 2014).     By contrast, at intermediate photon energies, knowledge of the spatial and temporal behaviour of $T$ is required, as are solutions of the relativistic evolution equation (\ref{fund}).  Approximate analytic solutions of the latter are discussed below. 

\subsection{Self-similar disc solutions}\label{discsolutions}
As a second order differential equation, equation (\ref{fund}) has two independent solutions. 
MB1 showed that the gross properties of relativistic discs can be understood by self-similar solutions of the simpler, large $r$, Newtonian evolution equation.   The key point is that the inner boundary condition, which is determined by the ISCO stress, directly affects the behaviour of the disc in the outer  zone (MB2). 

Self-similar solutions exist when the turbulent stress is of the form
\beq
\W = \omega \left(\frac{\Sigma}{\Sigma_0}\right)^\eta \left(\frac{r}{r_0}\right)^\mu ,
\eeq 
which is applicable to a rather wide range of turbulent disc models, including classical $\alpha$-discs (Shakura \& Sunyaev 1973).  (The length $r_0$ and density $\Sigma_0$ are set by the initial  conditions.)  Introducing the dimensionless space and time variables 
\beq\label{taudef}
x = r/r_0, \quad \tau =  \omega t/\Big(2\sqrt{GMr_0^3}\Big) ,
\eeq
the solutions have the general form (Pringle 1991):
\begin{align}\label{discsolutions}
\Sigma &= \Sigma_0 \,p(y)\,x^{-3/2} \,\tau^{-\chi}, \\
 p(y) &= y^C (1-ky^B)^{1/\eta}, \\
 y &= {x}^{1/2}\, \tau^{-\lambda} .\label{discsolutions2}
\end{align}
For a given $\eta$ and $\mu$,  the two independent solutions have this same general form, differing only in their constant parameters. We have verified by direct numerical integration of equation (\ref{fund}) that over the extended timescales of interest, the exact solution is generally dominated by one or the other of the self-similar profiles (MB1), the choice being dictated by the nature of the ISCO stress.

The coefficients may be expressed in terms of $\eta$ and $\mu$ (Pringle 1991). For a finite ISCO stress  the coefficients are  
\begin{align}
B &= (5\eta + 3 - 2\mu)/(\eta + 1),\\
C &= (3\eta + 1 - 2\mu)/(\eta + 1), \\
k &= \eta/[ (4\eta + 3-2\mu)(5\eta+3-2\mu)], \\
\lambda &= 1/(4\eta + 3 - 2\mu), \\
\chi &= 1/(4\eta + 3 - 2\mu) ,
\end{align}
whereas for a vanishing ISCO stress they are given by
\begin{align}
B &= (4\eta + 3 - 2\mu)/(\eta + 1), \\
C &= (3\eta + 2 - 2\mu)/(\eta + 1),  \\
k &= \eta/[(4\eta + 3 - 2\mu)(5\eta + 3 - 2\mu)],  \\
\lambda &= 1/(5\eta + 3 - 2\mu),  \\
\chi &= 2/(5\eta + 3 - 2\mu). 
\end{align}
Substituting the Newtonian self-similar solutions (\ref{discsolutions}-\ref{discsolutions2}) into the temperature equation (\ref{temperature}) leads, after algebraic simplification, to a disc temperature profile of the form
\beq\label{disctemp}
T(x,\tau) = T_0 \  x^{-m} \, \tau^{-n/4} \, \psi(x,\tau) , 
\eeq
where 
\beq\label{tempscale}
T_0 = \left(\frac{3 \omega \Sigma_0}{4 \sigma} \sqrt{\frac{GM}{r_0^5}}\right)^{1/4} ,
\eeq
and
\beq
\psi(x,\tau) = \left(1 - k\, x^{B/2} \tau^{-\lambda B} \right)^{(1+\eta)/4\eta}.
\eeq
The power law indices $m$ and $n$ are given by, for a finite ISCO stress
\begin{align}
m &= 7/8 , \\
n &= \frac{4\eta+2-2\mu}{4\eta +3 - 2\mu} , \label{nf}
\end{align}
and for a vanishing stress 
\begin{align}
  m &= 3/4 , \\
n &=  \frac{5\eta+4-2\mu}{5\eta +3 - 2\mu} . \label{nv}
\end{align}
The index $n$ is the same index that appears in the bolometric luminosity $L \sim t^{-n}$.   While $n$ depends only rather weakly on $\eta$ and $\mu$, it depends more sensitively upon the ISCO stress:  $n \approx 0.8$ for finite ISCO stress modes versus $n \approx 1.2$ for vanishing ISCO stress modes (MB1).  
When confusion may arise from the different values of `$n$', we shall denote the finite ISCO stress decay index (eq. \ref{nf}) by $n_f$, and the vanishing ISCO stress decay index by $n_v$ (eq. \ref{nv}). 

\subsection{Rayleigh-Jeans limit}
The Rayleigh-Jeans limit describes the lowest frequency part of the spectrum, where $h\nu_\oo \ll k_B \widetilde T$ for all effective temperatures within the disc.  At these frequencies we may expand the exponential:
\beq
\exp\left({h\nu_\oo}/{k_B \widetilde T} \right) - 1 = {h\nu_\oo}/{k_B \widetilde T} \,+ \, ... \,,
\eeq
and thus the time dependence of the Rayleigh-Jeans flux follows that of the disc temperature.    At late times we obtain from equations (\ref{flux2} \& \ref{disctemp})
\beq
F_\nu(\nu_\oo,t) \sim \nu_\oo^2 \, t^{-n/4} .
\eeq

\subsection{Mid-range frequencies}\label{midrange}
Each disc radius has both a characteristic temperature (\ref{disctemp}) and charateristic frequency $\nu \sim k_B T/ h$. At intermediate frequencies where $T_{\rm out} \ll h\nu/k_B \ll T_0$, where $T_{\rm out}$ is a typical temperature of the  outer disc,  the flux originates primarily from (Newtonian) regions where the self-similar solutions (\ref{discsolutions}) are a good approximation.   In this regime, relativistic Doppler shifts are negligible ($f \simeq 1$) and gravitational optics unimportant: photons travel  in nearly straight lines.  This allows a further simplification of the integrals by using the standard Newtonian form for the solid angle element of emission, 
\beq
 \text{d}\Theta_{{o}} = \frac{\text{d}\alpha \, \text{d} \beta}{D^2} = \frac{2\pi r }{D^2}  \cos\theta_{\text{obs}} \, \text{d}r \, .
\eeq
In this limit, the flux can be written as an integral over the dimensionless disc radius $x=r/r_0$ from a suitable inner radius $x_{\rm in}$ to the outer radius $x_{\rm out}$:
\begin{equation}\label{FSintegral}
F_\nu(\nu_o,t) =A \int_{ x_\text{in}}^{ x_{\text{out}}} x  \left[ \exp \left(\frac{\epsilon x^{m} \tau^{n/4}}{\psi(x,\tau)} \right) - 1\right]^{-1} \, \text{d}x \, ,
\end{equation}
where the constants $A$ and $\epsilon$ are given by
\beq
A = 4\pi h r_0^2 \nu_o^3\cos\theta_{\rm obs}/(c^2 D^2), \quad \epsilon = h\nu_o/k_BT_0.
\eeq
At late times, the dominant contribution of the mid-frequency flux comes from radii where $x^{1/2} \,\tau^{-\lambda}$ is small and $\psi\simeq 1$.   With little contribution coming from either large or small radii, to leading order the limits of integration can be extended  from $0$ to $\infty$, and the integral becomes 
\beq
F_\nu = A \int_0^\infty x \left[ e^{ \epsilon x^{m} \tau^{n/4} } - 1\right]^{-1} \, \text{d}x. 
\eeq
This may be evaluated explicitly:
\beq
F_\nu = {qA\over 2}\ \Gamma(q)\,  \zeta_{\cal R} (q)\,  \epsilon^{-q} \tau^{-nq/4} \sim \nu_o^{3- q} \, t^{-nq/4} .
\eeq
where $q=2/m$, and $\Gamma$ and $\zeta_{\cal R}$ are the standard Gamma and Riemann zeta functions respectively (cf. Lynden-Bell 1969).     
For the finite ISCO stress mode $m = 7/8$, and the observed flux is 
\beq\label{flatsection}
F_\nu(\nu_o,t)= 1.9039 \,A\epsilon^{-16/7} \tau^{-4n_f/7}  \sim \nu_o^{5/7} t^{-4n_f/7} ,
\eeq
whereas for the case of vanishing ISCO stress we find
\beq\label{flatsec2}
F_\nu(\nu_o,t)= 2.5762 \, A\epsilon^{-8/3} \tau^{-2n_v/3}  \sim \nu_o^{1/3} t^{-2n_v/3}.  
\eeq
The frequency dependence $F_\nu \sim \nu^{5/7}$ that is characteristic of a finite ISCO stress arises directly from the modified $T \sim r^{-7/8}$ temperature profile (cf.\ Agol \& Krolik 2000).    For a vanishing stress disc  $T \sim r^{-3/4}$ and we recover the classic Lynden-Bell (1969) $F_\nu\sim \nu^{1/3}$ scaling. 

\subsection{Evolution of the mid-frequency spectral shape}\label{spectralshape}
Our results suggest that when observed at frequencies below the thermal spectrum peak, some care is required when extracting the exact time dependence of the observed flux in a narrow observational band.  As the disc cools and expands, the spectrum peaks at lower frequencies and decreases in overall magnitude.   The peak frequency $\nu_p$ is related to the peak effective temperature by 
\beq\label{freqpeak}
\nu_p \sim \frac{k_B \widetilde T_p}{h} \sim t^{-n/4} .
\eeq
(We have used eq.\ [{\ref{disctemp}] for the time dependence of $\widetilde T_p$; see \S \ref{QWlim} below for further discussion.)
Below the peak of the spectrum, the flux increases as a power law with $\nu$ (eqs. [\ref{flatsection}] and [\ref{flatsec2}]).   The shifting of the peak of the spectrum to lower frequencies combined with a decreasing overall magnitude of the spectrum means that the observed flux at a particular frequency (or in a surrounding narrow band) will tend to remain nearly constant---or even increase---with time.   Only at much later times will the observed flux follow the power-law declines derived earlier [equations (\ref{flatsection}) and (\ref{flatsec2})].  As we shall see, this subtle behaviour appears to present in the observed  narrow band light curves of ASASSN-14li. 

\subsection{Quasi-Wien limit}\label{QWlim}
The quasi-Wien limit of the disc spectrum is characterised by the largest photon energies,
\beq\label{Wiendef}
h \nu_o \gg k_B \widetilde T_p .
\eeq
In this limit,  all of the emission is strongly concentrated near the temperature maximum, and we may simplify the denominator of 
(\ref{flux2}) by neglecting the $-1$, since this introduces only exponentially small corrections.  The flux is then formally given by 
\beq\label{flux3}
F_\nu(\nu_o,t) =  {\frac{2h\nu_o^3}{D^2c^2}}\iint_{\cal S}  \exp\left(- {h\nu_o}/{k_B \widetilde T} \right) ~ {\text{d}\alpha \, \text{d} \beta} .
\eeq
Since only the hottest parts of the disc very near the temperature maximum contribute to the observed flux, the integral may be evaluated using Laplace's method (e.g.\ Bender \& Orszag 1978).   We follow Balbus (2014) by Taylor expanding in the inverse of the effective temperature about the effective temperature maximum.  

The details will depend both on the properties of the ISCO stress  (which affect the location of the disc temperature maximum), and the disc viewing angle (which affects the projected geometry of the hot inner disc).   Current models of TDE evolution suggest that there should be a rather strong viewing angle dependence of the observed flux, with observable X-ray emission only when the disc is orientated near face-on (Dai \textit{et al}. 2018).  We shall therefore assume here that the disc is oriented nearly face-on.   Asymptotic solutions of equation (\ref{flux3}) may be straightforwardly obtained for different disc orientations and ISCO stress assumptions;  these are presented in \S\ref{FvsVstress} (vanishing ISCO stress), and Appendix B (different orientation angles).

For a face-on disc, there is little kinematic Doppler shifting of the observed radiation, and the effective temperature profile is therefore azimuthally symmetric in the image plane.   We may therefore expand in the radial image plane coordinate $R$:
\beq 
R \equiv \sqrt{\alpha^2 + \beta^2} .
\eeq
The expansion of the effective temperature is then 
\beq\label{Teffexpansion}
\frac{1}{k_B \widetilde T} = \frac{1}{k_B \widetilde T_p} + \frac{1}{k_B \widetilde T_p^2} \left| \frac{\partial \widetilde T}{\partial R} \right|_{R_p} \left(R- R_p\right) + ...,
\eeq
where we have used the fact that for a disc with a finite ISCO stress the peak temperature $\widetilde T_p$ lies on the inner disc boundary at radial coordinate $R_p$ in the image plane.   For clarity, the (negative) temperature gradient appears here as an absolute value.   It is possible to expand the observed flux to arbitrarily high powers of the small parameter 
\beq
\delta \equiv \frac{k_B \widetilde T_p}{h \nu_o} \ll 1 ,
\eeq
by keeping higher order terms in the Taylor expansion (\ref{Teffexpansion}).   We will see, however, that for our purposes the leading order behaviour is sufficient.  While all of the spatial dependence of the effective temperature is explicit in equation (\ref{Teffexpansion}), it is important to remember that the peak effective temperature and effective temperature gradient are time-dependent quantities. Substitution of (\ref{Teffexpansion}) into (\ref{flux3}) leaves 
\begin{multline}\label{flux4}
F_\nu(\nu_o,t) =  {\frac{4\pi h\nu_o^3}{D^2 c^2}}\exp\left(- \frac{h\nu_o}{k_B \widetilde T_p} \right)   \times \\
\int_{R_p}^\infty R \,  \exp\left(- \frac{h\nu_o}{k_B \widetilde T_p^2} \left| \frac{\partial \widetilde T}{\partial R} \right|_{R_p} \left(R- R_p\right) \right)  \, \text{d}R . 
\end{multline} 
The integral is now elementary.  
We define the order unity numerical constant $\xi$ by
\beq\label{xidef}
 \xi  \equiv  \frac{R_p}{\widetilde T_p} \left| \frac{\partial \widetilde T}{\partial R} \right|_{R_p}.  
\eeq
This enters the expression for the flux as a normalisation factor.  (For a finite ISCO stress Newtonian disc, $\xi =7/8$.)  The resulting high frequency spectrum is
\beq
F_\nu(\nu_o,t) = \frac{4\pi}{\xi c^2} \left(\frac{R_p}{D}\right)^2  \left(\frac{k_B \widetilde T_p }{h \nu_o} \right) h \nu_o^3 
\exp\left(- \frac{h\nu_o}{k_B \widetilde T_p }  \right). 
\eeq
We recover the modified-Wien $\nu^2$ frequency dependence first noted by Balbus (2014). The implicit time dependence of the flux enters via the disc temperature.  The integral over the X-ray frequency band then follows straightforwardly:  
\beq\label{FXint}
F_X(t) = \int_{\nu_l}^\infty F_\nu(\nu_o,t) \,\text{d}\nu_o ,
\eeq
where the lower limit $\nu_l$ corresponds to $0.3$ keV for the \textit{Swift} ASASSN-14li observations, and the upper integration limit is set to infinity (introducing only exponentially small corrections). This gives to lowest order
\beq
F_X(t) =  \frac{4\pi}{\xi c^2} \left(\frac{R_p}{D}\right)^2 \left( \frac{k_B \widetilde T_p}{h \nu_l} \right)^2 h\nu_l^4 \, 
 \exp\left(- \frac{h\nu_l}{k_B \widetilde T_p}  \right)  ,
\eeq
which is quadratic in the small parameter $\delta$.
\begin{figure}
 \centering
  \includegraphics[width=0.5\textwidth]{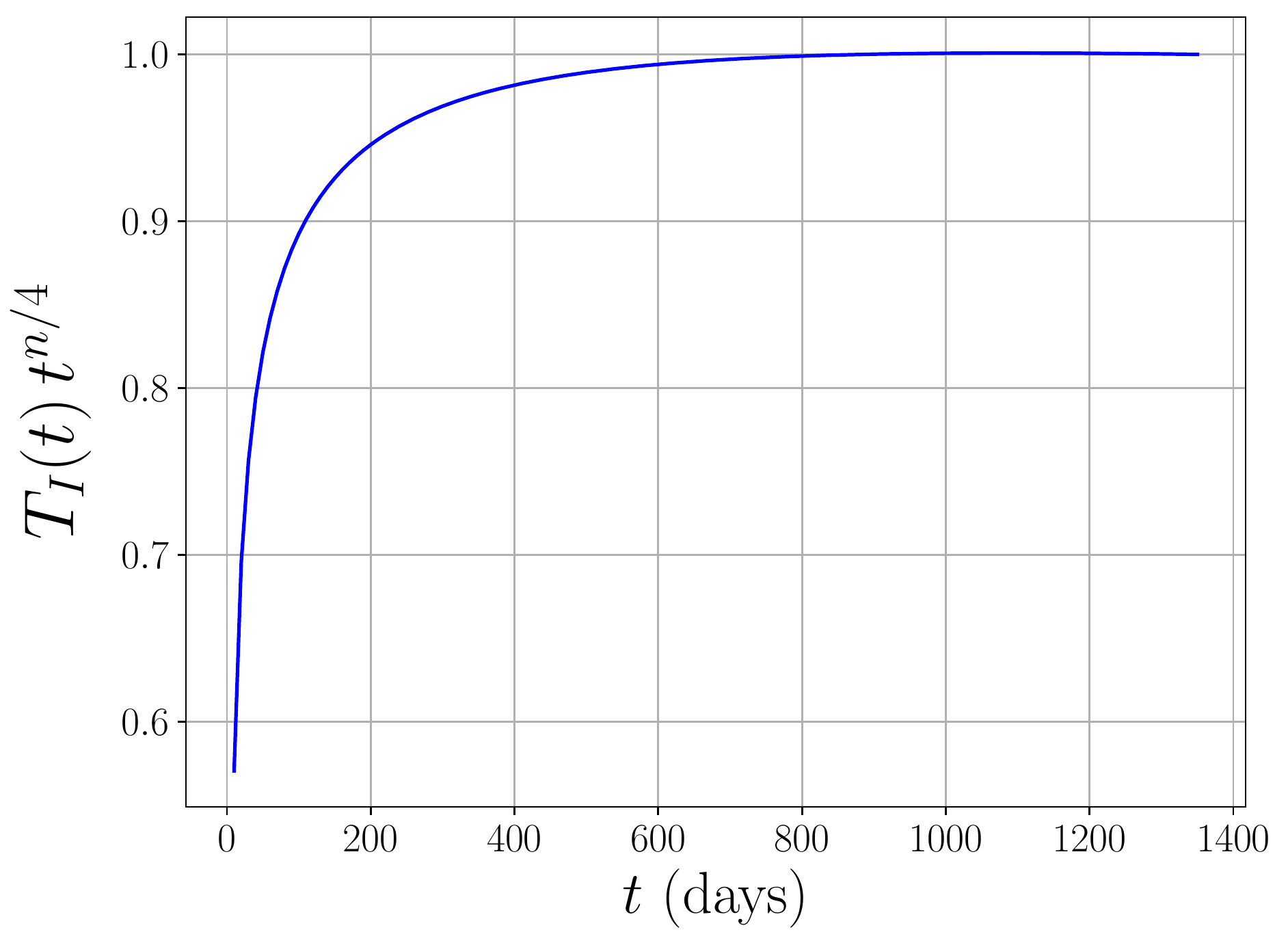} 
 \caption{ISCO temperature versus time.  The approximation of equation (\ref{Tdef}) is valid to within a few percent after 200 days. The y-axis is normalised so as to equal 1 at a time corresponding to the end of the available observations of \Asa. This plot was produced with $n = 0.86$. }
 \label{ISCOtemp}
\end{figure}

We next revisit the question raised in \S 3.5, namely what form does the time dependence of the peak disc temperature take?  With $x_p$ fixed, in a Newtonian disc the peak temperature follows a simple power law in time $T_p \sim t^{-n/4}$ (equation \ref{disctemp}).  In a relativistic disc, the temperature in the innermost regions is not in general  given by equation (\ref{disctemp}).  However, the temporal behaviour of the disc temperature in the innermost regions is unlikely to be systematically different from that of the outer disc, since the Newtonian and relativistic regimes of the disc must join smoothly at all times.   We therefore assume a peak temperature functional form of
\beq\label{Tdef}
\widetilde T_p(t) \equiv \T_p \,\tau^{-n/4} ,
\eeq
where $\T_p$ is a constant with dimensions of temperature and $\tau$ is the time in units of the viscous time scale. The viscous timescale $\tau$ {for a Green's function solution is defined in terms of the initial radius $r_0$ (eq. \ref{taudef}), and} could in principal include a time offset $t_0$ to provide a finite X-ray luminosity as $t\rightarrow 0$. This assumption has been checked and explicitly verified by exact numerical calculation {(Figure \ref{ISCOtemp}).}  This result now allows us to calculate the explicit time dependence of $F_\nu$ and $F_X$:
\begin{multline}\label{flux5}
F_\nu(\nu_o,t) = \frac{4\pi}{\xi c^2} \left(\frac{R_p}{D}\right)^2 \left(\frac{k_B \T_p}{h \nu_o} \right) h\nu_o^3\  \times   \\ \tau^{-n/4}
\exp\left(- \frac{h\nu_o}{k_B \T_p} \tau^{n/4} \right) ,
\end{multline}
and 
\begin{multline}\label{FX3}
F_X(t) = \frac{4\pi}{\xi c^2} \left(\frac{R_p}{D}\right)^2 \left( \frac{k_B \T_p}{h \nu_l} \right)^2 h\nu_l^4 \  \times \\
\tau^{-n/2} \, \exp\left(- \frac{h\nu_l}{k_B \T_p} \tau^{n/4} \right) ,
\end{multline}
with $n \approx 0.8$.    The underlying assumptions and approximations that have gone into equation (\ref{FX3}) are:
\begin{enumerate}
\item{The emission is well approximated as thermal (eq. \ref{flux}).}
\item{The emission is in the quasi-Wien limit, $h\nu_l \gg k_B \T_p$.}
\item{The peak disc temperature follows $T_p \sim t^{-n/4}$.}
\item{The peak temperature occurs at the innermost disc edge.}
\item{The disc is viewed close to face-on.}
\end{enumerate} 
{Note that, for the solutions under consideration, the assumption that the peak temperature lies at the inner disc edge is, in effect, equivalent to assuming a finite ISCO stress. }
These results are likely to be valid under conditions rather more general than the thin disc model {\it per se}.  In Appendix \ref{appendixC} we present analytic expressions for the time dependent X-ray emission, relaxing the final  approximation.  The functional form of (\ref{FX3}) turns out to be rather insensitive to the precise viewing angle. 

\subsection{Summary}
\begin{figure}
 \centering
  \includegraphics[width=0.5\textwidth]{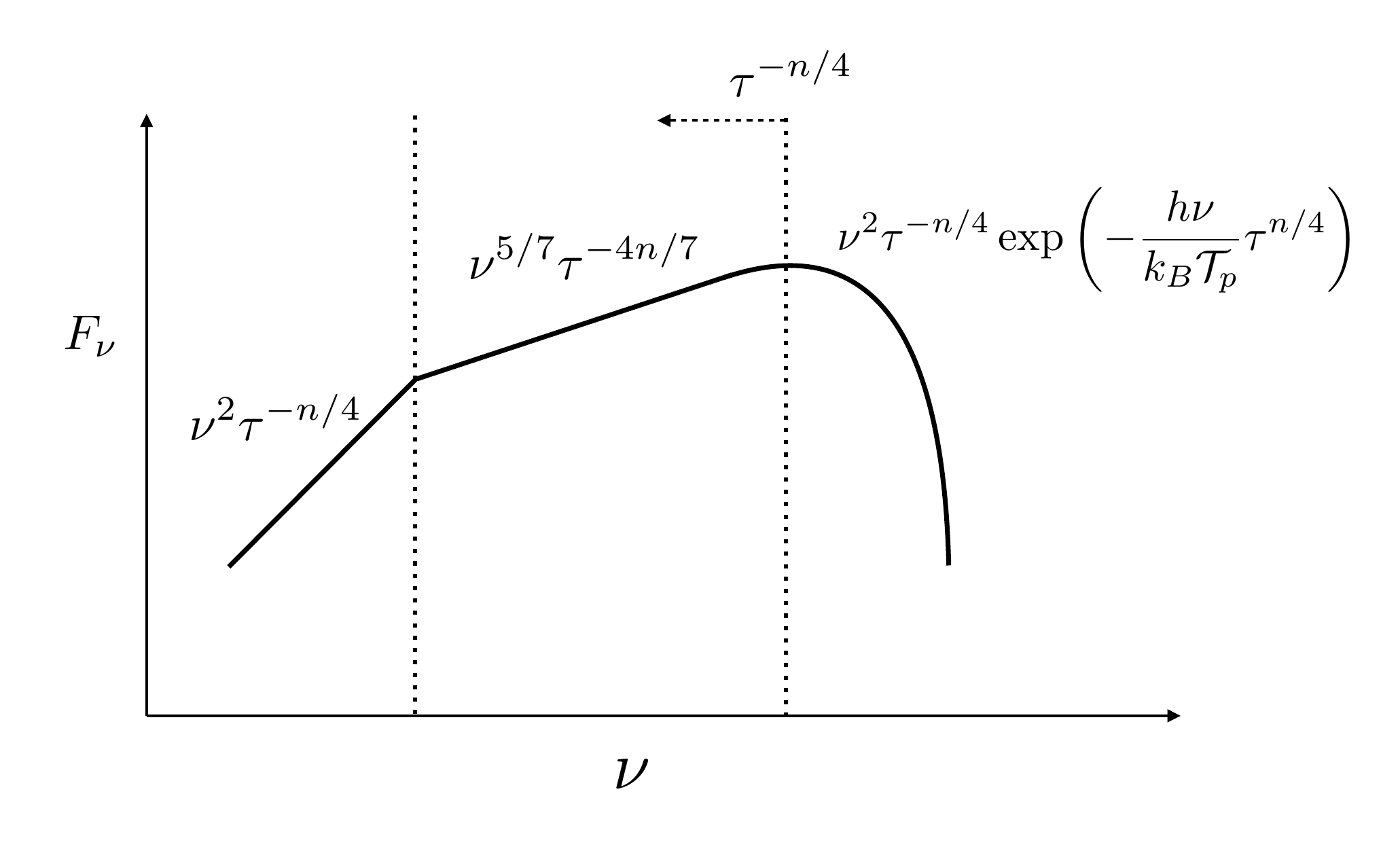} 
 \caption{A schematic finite ISCO stress disc spectrum.   The low, mid, and high frequency asymptotic regimes are delineated.  (The effects of the evolving boundaries between the frequency domains may be seen in the numerical solutions displayed in Fig.\ [\ref{evolvingspectrum}].)  }
 \label{schematic}
\end{figure}
The observed flux in several diffrent limits of physical interest may be obtained by taking appropriate expansions of the integral (\ref{flux}) {[Figure \ref{schematic}]. }
We find that that even if the bolometric luminosity follows a simple power law decline
$L \sim t^{-n}$, 
the observed finite bandpass flux will often have a different temporal behaviour.    For example, if X-ray observations probe the quasi-Wien tail of the evolving disc spectrum, then rather than a power law decay,  the X-ray light curves of transient disc sources is described by a function of the form 
\beq
F_X(t) = F_0 \left(\frac{t + t_0}{t_X}\right)^{-n/2} \exp\left[- \left(\frac{t + t_0}{t_X}\right)^{n/4}\right] ,
\eeq
with $n \approx 0.8$, and $F_0, t_0$ and $t_X$ time-independent constants. At the very largest times the X-ray light curve becomes dominated by an exponential of a power law in time, with power law index less than 1 in magnitude.   We therefore expect the true X-ray luminosity of a TDE source to follow a decline that is steeper than a power-law but more shallow than a pure exponential.  
Furthermore, if the UV luminosity is dominated by disc emission, this bandpass should be characterised by a nearly constant luminosity.  This (perhaps) counterintuitive results stems from the combination of a decreasing temperature coupled with an increasing radial extent that places the peak of emission at smaller frequencies.   At frequencies below but near the peak frequency of the disc spectrum, the observed flux remains near constant.   These results will be discussed in detail in \S 4.   

\section{ASASSN-14li numerical disc model}\label{numerical}
\subsection{Observations}

How well can a thin disc model explain the spectral observations of the confirmed TDE, ASASSN-14li (Holoein \textit{et al}. 2016),  at both far UV and X-ray wavelengths?  The X-ray observations used here are taken from Bright \textit{et al}. (2018), and are publicly available.  The UV observations, predominantly taken from Brown \textit{et al}. (2017), also use late time observations from Van Velzen \textit{et al}. (2019).   The UV observations span $\sim 1300$ days, the X-rays $\sim 900$ days.   The observations are host-subtracted:  we assume that the observed flux is intrinsic flux from the TDE source. 

The technique is straightforward.  One specifies a set of black hole properties ($M$, $a$) and disc parameters [$\Sigma(t=0)$, $\W$], and solves the evolution equation (\ref{fund}) for the time-dependent disc temperature (\ref{temperature}).     For a given disc inclination angle $\mu = \cos(\theta_{\text{obs}})$,  ray tracing calculations allow the observed redshift (\ref{redshift}) to be calculated.  
We calculate spectra at (narrow band) UV and (broad band) X-ray wavelengths.   Such simultaneous fits at frequencies which differ by two orders of magnitude tightly constrain the model parameters. 

The X-ray flux is given by (eq. \ref{FX}).  For the narrower UV bands, we simply compute the evolving flux at the central wavelength of each UV band.  For the three \textit{Swift} bands of interest (UVW1, UVW2 and UVM2) these are\footnote{http://www.swift.ac.uk/analysis/uvot/filters.php} :
\begin{align}
\bar \lambda_{W1} &= 260.0 \, \text{nm} , \\
\bar \lambda_{M2} &= 224.6  \, \text{nm}, \\
\bar \lambda_{W2} &= 192.8  \, \text{nm}.
\end{align}
The UV results are presented as evolving AB magnitudes, where 
\beq\label{magdef}
m_{AB}(\bar\lambda,t) \equiv - 2.5 \log_{10}\left[F_\nu\left(c/\bar \lambda,t\right)\right] - 48.60.
\eeq
Here, $F_\nu$ is calculated from equation (\ref{flux}), and has units of erg/s/cm${}^2$/Hz. This approach ignores any effects from the actual finite width of the UV bands; these are likely to be unimportant for present purposes.    The UV bands are displayed as vertical lines in Fig.\  (\ref{examplespectrum}) against our disc spectrum, computed 200 days after disc formation using the best fit parameters.   

\subsection{Model parameters} 
In general, there are nine parameters which completely specify an evolving disc model. These parameters, which are described below,  involve the observer orientation, the initial disc conditions, the turbulent stress, and the mass and angular momentum of the central black hole.  
\subsubsection{Observer parameters}
This distance to the source serves as an overall normalisation of the observed flux, one that often is not accurately known. The distance to ASASSN-14li is the exception however: it is tightly constrained by its well identified host galaxy: PGC 043234 (Holoein \textit{et al}. 2016), at a distance $D =90$ Mpc.  The disc-observer inclination angle is unknown for ASASSN-14li.   As earlier noted, current TDE models suggest that sources observed at X-ray wavelengths are expected to be orientated near face-on, $\theta_{\text{obs}} \lesssim 20^\circ$ (Dai \textit{et al}. 2018).  Given the lack of observational constraints and the expectation of a near face-on orientation, we shall assume for simplicity that 
$ \cos ( \theta_{\text{obs}} ) = 1$.
In principle,  our approach could be readily extended to model the observed X-ray and UV {\bf fluxes} for a variety of different viewing angles. 

\subsubsection{Central black hole parameters}
A stationary black hole is completely described by exactly two parameters: its mass and spin. 
The mass of the central black hole is somewhat constrained by previous observations.  Using measurements of the host galaxy's bulge mass and standard SMBH-host galaxy scaling relationships,  Mendel \textit{et al}. (2014) derived a black hole mass of $M \sim 10^{6.7 \pm 0.6} M_{\sun}$. {Somewhat tighter constraints were found by Wevers {\it et al}. (2017), who performed velocity dispersion measurements and found a black hole mass of $M = 10^{6.23 \pm 0.4} M_{\odot}$.   } On the other hand,  using Eddington-limit arguments Miller \textit{et al}. (2015) inferred a mass of $M \sim 2 \times 10^6 M_{\sun}$.   As these relationships are broadly uncertain, in our model fitting we have allowed the black hole mass to vary within the range 
\beq
10^{6} < M /M_{\sun} < 10^7 .
\eeq

The angular momentum of the central black hole is even less tightly constrained by observations. Pasham  {\it et al}. (2019) reported a recent measurement of quasi-periodic oscillations (QPOs) in the observed ASASSN-14li X-ray spectrum. This was used to set upper and lower bounds on the black hole parameters: a high prograde spin ($a/r_g > 0.7$), and a mass less than $M = 2 \times 10^6 M_{\sun}$.    However, the exact black hole mass and spin parameters inferred from QPO measurements are heavily dependent upon the underlying model for the origin of the QPO, something that is at present uncertain.   (Pasham {\it et al.} assume an orbiting test particle model).  Here, we shall present our own analysis of the angular momentum of the ASASSN-14li black hole.   In the following section we model the spectral fits for a somewhat more slowly rotating Kerr black hole, $a/r_g = 0.5$. 
In \S\ref{BHproperties} we study the effect of varying the black holes spin on the evolving disc light curves.

\subsubsection{Turbulent stress modelling} 
As in our previous studies, we consider a turbulent stress of the general form
\beq
\W = \omega \left(\frac{\Sigma}{\Sigma_0}\right)^\eta \left(\frac{r}{r_0}\right)^\mu .
\eeq
The precise parameterisation of $\W$ affects the time dependent properties of the disc in two ways. The first is through its effects on the bolometric decay index $n$.   This enters several expressions for the time dependent flux at a number of different wavelengths (e.g, equations \ref{flatsection} \& \ref{FX3}).  The value of $n$ depends only weakly on the exact parameterisation of the stress (the indices $\eta \,\& \,\mu$).   By contrast, it is strongly dependent on the nature of the stress at the ISCO. The bolometric decay index $n$ is given by 
\beq
n = \frac{4\eta+2-2\mu}{4\eta +3 - 2\mu} ,
\eeq  
for a finite ISCO stress (MB1), whereas for a vanishing ISCO stress  
\beq
n = \frac{5\eta+4-2\mu}{5\eta +3 - 2\mu} .
\eeq  
Because of the limited effect of the exact turbulent stress parameterisation on the decay index $n$, we employ a simple self-similar turbulent stress model $\W = w = $ constant (i.e, $\eta = \mu = 0$). This model is also physically rather natural:  it would emerge, for example, if the turbulent velocity fluctuations follow a Keplerian scaling of the form $\delta v^r \sim \delta v^\phi \sim r^{-1/2}$. 

We assume that the stress is \textit{finite} at the ISCO.  To avoid  unsustainable behaviour at lates times, we follow the `quasi-circular' orbit approach developed in MB2. This allows for deviations in the mean fluid motion of the disc from that of exact Kerr-circular orbits. These deviations must of course be present at some level, as a consequence of non-zero radial velocities in the innermost disc regions. The degree to which the mean fluid motion deviates from perfect circular orbits may be parameterised by a dimensionless number, denoted as $\gamma$ (MB2). 
{The $\gamma$ parameter is defined by 
\beq
\gamma \equiv \frac{\Delta j \, u^r}{\W},
\eeq
where $u^r$ and $\W$ are the radial velocity and turbulent stress measured at the ISCO, and $\Delta j$ is the difference in the disc fluids ISCO angular momentum from that of a circular orbit.   A finite ISCO stress disc model with precisely circular orbits would therefore have $\gamma = 0$, whereas a vanishing ISCO stress disc must have $\gamma \rightarrow \infty$. The properties of $\gamma$ cannot as yet be calculated from first principles.   We therefore must rely on simulations to determine its value and variability.    }
 Current estimates based upon GRMHD simulations place $\gamma$ within the range $0.01 < \gamma < 0.1$.  We adopt the midpoint of this range, $\gamma = 0.05$. The dimensionless parameter $\gamma$ sets the time scale over which the finite ISCO stress modes (eq. \ref{discsolutions}) dominate the behaviour of the disc (MB2).  For small $\gamma$, the disc evolution is dominated by these modes for the entire observational period.  

The next important parameter is the magnitude of the turbulent stress, $w$.  For an evolving disc, $w$ sets the duration of the ``viscous''  timescale $t_v$.   For our choice of $\eta$ and $\mu$ this is given by 
\beq\label{visctime}
t_v = 2\sqrt{GMr_0^3}\big/9w ,
\eeq
where $r_0$ is a radius of the the initial disc ring (Appendix A, BM18).  The stress amplitude $w$ is determined from a best fit to the viscous evolution timescale, calculated using the above.

\subsubsection{Disc initial condition}
The initial event which produced ASASSN-14li was not observed.   The source was first observed on 22 November 2014. Prior to this date, its host galaxy PGC 043234 had not been observed since the 13th of July 2014, when no TDE was present, as the sun was obscuring this galaxy. This leaves a  $132$ day window between observations within which the initial disruption could have occurred.  We therefore leave the `disc formation time' $t_D$ (the time at which we begin the disc evolution), as an additional free parameter:
\beq
-132 < t_D  - 56983.6 \, (\text{MJD}) < 0 ,
\eeq
where 56983.6 MJD corresponds to the Julian Date of the first recorded observation.

We take the initial density distribution of our disc to be a numerical delta function at a fixed radius of $r_0 = 15r_g$.   {This radius is equal to the tidal radius of a main sequence star with the mean stellar mass $M_\star = 0.36 M_{\odot}$ (Kroupa 2001) around a $2 \times 10^6 M_{\odot}$ black hole, the best estimate of the black hole mass prior to fitting. 

The process of disc formation post-TDE is dynamically complex.  In reality, the initial matter distribution will depend upon the structure and orbit of the pre-disruption star and the mass and spin of the central black hole. This initial condition is thus an idealisation.   However,}
the behaviour of disc light curves post-peak are in fact well-described by sets of self-similar solutions (eq. \ref{discsolutions}), and these have lost memory of the initial conditions (MB1).  Comparisons to the post-peak observations of ASASSN-14li are thus likely to be insensitive to the precise spatial form of the initial density distribution. 

Our final fitting parameter is the initial mass of the disc, calculated by
\beq
M_d = 2\pi \int\limits_{r_I}^{\infty} (g_{rr} g_{\phi\phi})^{1/2} \, \gamma^\phi(r,a) \, \Sigma(r, t=0) \, \text{d} r .
\eeq
Here $\gamma^\phi$ is a relativistic factor relating the disc area element in the rotating disc frame to that of the Boyer-Lindquist co-ordinate system (see Bardeen {\it et al}. [1972] for a discussion). 

\subsection{Best fit parameters}

\begin{table}
\renewcommand{\arraystretch}{2}
\centering
\begin{tabular}{|p{2.5cm}|p{2.1cm}|}
\hline
$M/ M_{\sun}$ & $1.85^{+0.03}_{-0.04} \times 10^{6}$ \\ \hline
$ M_d/ M_{\sun} $ & $ 1.63^{+0.07}_{-0.02} \times 10^{-2}$ \\ \hline
$t_D - 56983.6$ (MJD) & $-43.6^{+5}_{-7}$ \\ \hline 
$t_v$ (days) & $44.3^{+2.0}_{-2.4}$  \\ \hline 
$\bar \chi^2_{\rm min}$ & 6.19 \\ \hline
\end{tabular}
\caption{The best fit parameters for the fiducial disc model, {and the reduced chi-squared of the best fit X-ray light curve. For comparison, the reduced chi-squared of the best fit power-law fit is $\bar \chi^2_{\rm pl} =  13.99$, and $\bar \chi^2_{\rm exp} = 8.54$ for an exponential fit. The formally large reduced chi-squared values are discussed further in section \ref{xrlc}.   } }  
\label{table1}
\end{table}

\begin{table}
\renewcommand{\arraystretch}{2}
\centering
\begin{tabular}{|p{2.5cm}|p{1.5cm}|}
\hline
$a/r_g$ & $0.5$ \\ \hline
$r_0/r_g$ & $15$ \\ \hline
$\gamma$ & $0.05$ \\ \hline
$D/\text{Mpc}$ & $90$ \\ \hline
$\cos \left( \theta_{\text{obs}}\right)$ & $ 1$ \\ \hline
\end{tabular}
\caption{Parameters fixed during fitting. }
\label{table2}
\end{table}

Table \ref{table1} summarises the best fit disc parameters for the recorded observations of ASASSN-14li. Table \ref{table2} summarises the parameters that remained fixed for the fitting procedure.  The four fitted parameters may be divided into two categories: those that predominantly affect the evolutionary time scale of the system, and those that set the overarching scale of the emission. 

The viscous time scale of the evolving disc $t_v$ and the disc formation time $t_D$ are well constrained by the X-ray observations, which are strongly time varying. {Our best fit time offset $t_D$ is also consistent with  an independent estimate of the TDE start time from radio observations (Alexander {\it et al.} 2016), which must pre-date the disc formation time.}  The remaining two parameters, the disc mass $M_d$ and the black hole mass $M$, are then constrained by the simultaneous fitting of the observed magnitudes of the UV flux and X-ray luminosity. Fitting either band independently is relatively easy, but this leaves large degeneracies in the fitted parameters.    This degeneracy is broken only when the light curves are fitted {\em simultaneously.} {See Appendix \ref{appendixB} for further detail and discussion.}

It is extremely encouraging that our best fit black hole mass  $M = 1.85 \times 10^6 M_{\sun}$ is consistent with a number of existing estimates ($M \sim 2 \times 10^6 M_{\sun}$ Miller \textit{et al}. [2015] , $M \lesssim 2 \times 10^6 M_{\sun}$ Pasham \textit{et al}. [2019], {$M = 10^{6.23 \pm 0.4} M_{\odot}$ Wevers {\it et al}. [2017] }and $M = 10^{6.7 \pm 0.6} M_{\sun}$ Mendel \textit{et al}. [2014]).  These estimates comprise a wide range of very distinct methods of evaluating the black hole mass.  

The best fit disc mass, at $M_d = 1.63 \times 10^{-2} M_{\sun}$, appears at first to be rather small.   Rees (1988) suggests $M_d \sim M_\star / 2$, where $M_\star$ is the mass of the star prior to disruption. {However, our best fit disc mass is in fact consistent with existing estimates of the total {\it accreted} mass of a number of TDE candidates. Holoein {\it et al.} (2016) estimated the total accreted mass of three TDEs, using the observed luminosity and simple energetic arguments. They found accreted masses in the range $M_{\rm acc} \sim 10^{-3}$ -- $10^{-2} M_{\odot}$. Given the uncertainties involved in estimating accreted masses, this is a broadly supportive finding.    }

 Furthermore, it is important to distinguish the inferred disc mass of our model with the total mass fraction of the debris which remains gravitationally bound in the aftermath of the TDE. Our inferred disc mass is a lower bound of the total debris mass, as it represents only the total evolving mass required to reproduce both the 900 days of observed X-ray emission and the late time UV emission. 
The observed UV and X-ray light curves are by no means the only emission stemming from ASASSN-14li.  For example, ASASSN-14li is also a source of significant radio emission (van Velzen {\it et al. 2016}, Bright \textit{et al}. 2018), which presumably stems from the presence of a jet.  An additional, rapidly decaying, light curve component is also required at early times to match the observations of ASASSN-14li at UV wavelengths (this will be discussed further in section \ref{fuvsec}).  Neither the mass content of the jet or this additional UV component contribute to our best fit disc mass.   Hence, the actual debris mass is likely to be significantly larger than our deduced disc mass.  

{The small disc-to-debris mass ratio can be naturally explained if, preceding its first observation, \Asa\ was accreting at super-Eddington rates.  The likely outflow could then produce a large and transient early time UV flux, whilst removing much of the debris mass, leaving a more tightly bound, low mass disc behind.   In this case, we would expect the bolometric luminosity of our disc to be near the Eddington limit at times close to the first observation.  This prediction will be examined further in section \ref{bol}. }

Finally, it is difficult to determine or constrain properties of the initial disc distribution, absent early time observations.   
For example, the temporal peak of our best fit X-ray light curve is chosen to coincide with the peak of the observed X-ray light curve (Fig. \ref{xray}).   This a choice of convenience, and yields a lower bound to the disc mass.   The uncertainty here is that if the X-ray flare were significantly more luminous at times preceding the first observation, the best fit mass could be significantly larger (so as to increase the inner disc temperatures and emitted X-ray flux).     

The dynamical physics of post TDE disc formation are still poorly understood.   Our result suggests that only a small fraction of the stellar material settles into an evolving disc at radii close to the black hole.  The fate of the majority of the stellar mass thus remains an interesting and open question.

\subsection{Disc evolution}
 \begin{figure}
 \centering
  \includegraphics[width=0.5\textwidth]{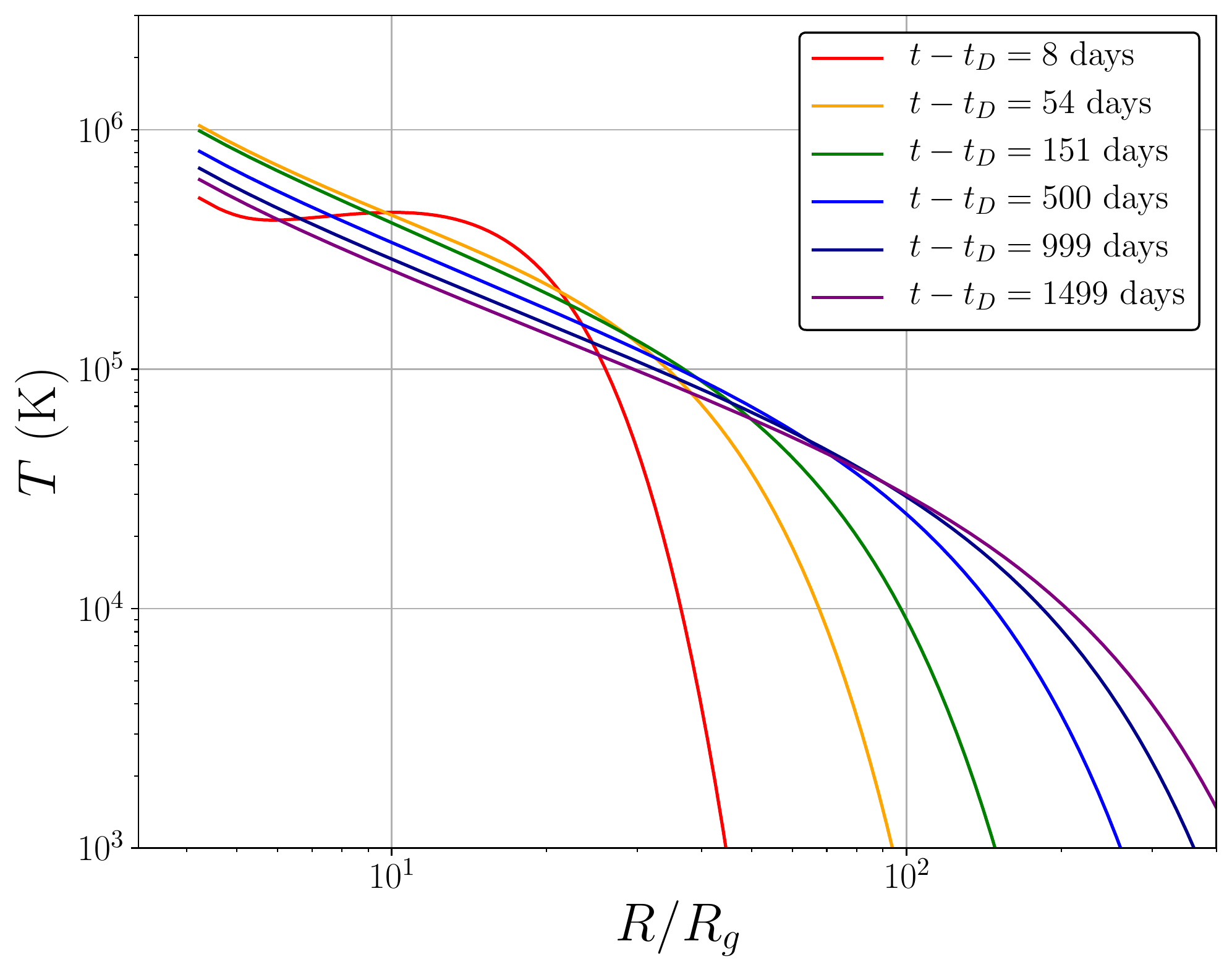} 
 \caption{Snapshots of the disc temperature profile at a number of different times (denoted on plot) for our best fit model. }
 \label{tempevol}
\end{figure}

{The evolution of the disc is qualitatively rather simple, as seen in Figure \ref{tempevol}.  The temperature in the innermost regions reaches its evolutionary maximum on a simple viscous timescale.   This is due to disc material rapidly spreading inwards towards the ISCO.  Once the disc density reaches a maxima in the inner regions the disc enters the `stalled-phase' of accretion (MB2).  This evolutionary phase is characterised by slow evolution of the inner disc (compare the temperature profile in the inner disc regions at 54 and 151 days), whilst the outer disc expands, taking up the angular momentum of the accreted material.   At yet later times, the disc begins its approach to the steady state:  the disc temperature gradually falls in the inner regions, and the outer region of the disc spreads to ever larger radii.    }

\subsection{Continuum spectrum}

 \begin{figure}
 \centering
  \includegraphics[width=0.5\textwidth]{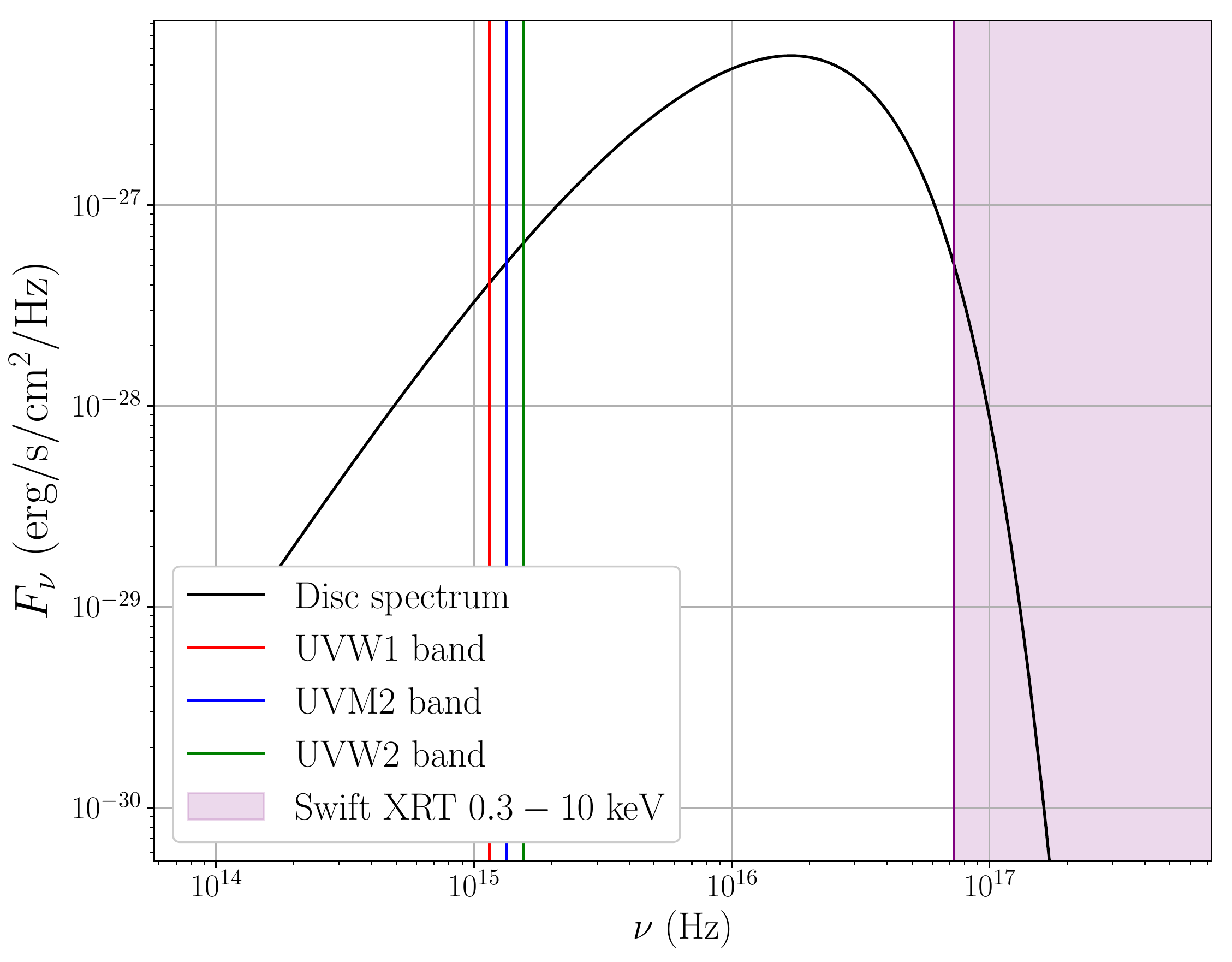} 
 \caption{Best fit disc spectrum 200 days after disc formation, the location of UV and X-ray bands is also shown.}
 \label{examplespectrum}
\end{figure}

 \begin{figure*}
 \centering
  \includegraphics[width=.75\textwidth]{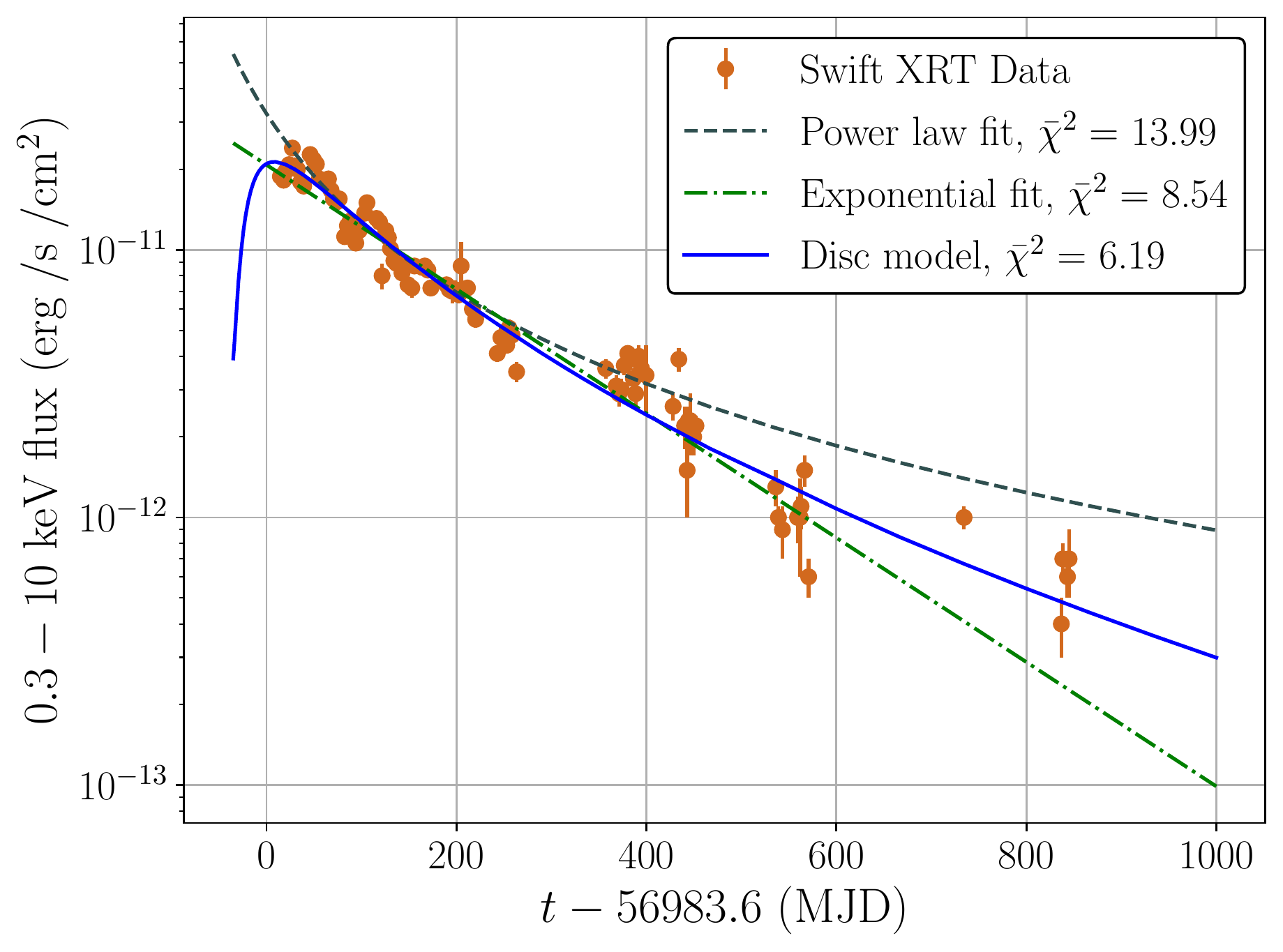} 
 \caption{Fiducial disc model fit (solid blue curve) to {\it Swift} X-ray observations of ASASSN-14li (Bright {\it et al}. 2018). Also plotted are the two best fit models from the literature, a power-law (grey dashed curve) and exponential fit (green dot-dashed curve; see text). The disc model has the lowest reduced chi-squared statistic,  providing the best fit to the observed \Asa\ light curve. }
 \label{xray}
\end{figure*}

{Figure (\ref{examplespectrum})} is a fiducial continuum spectrum 200 days after disc formation. This was calculated using equation (\ref{flux}) and the best-fit parameters of Tables \ref{table1} and \ref{table2}.  The three central frequencies of the UV bands are shown as coloured lines,  and the \textit{Swift} X-ray telescope band is the dashed area on the right of the figure.  This plot demonstrates that the observed UV flux is below the peak frequency, where we would expect the evolving flux to be near constant with time.   The {\it Swift} X-ray bands probe the quasi-Wien tail of the spectrum, and should therefore be described by equation (\ref{FX3}).

\subsection{X-ray light curve}\label{xrlc}

{Figure (\ref{xray}) } shows the observed ASASSN-14li $0.3 - 10$ keV X-ray flux (Bright \textit{et al} 2018), together with the best fit evolving X-ray flux from our fiducial disc model, and previous functional fits from the literature  (Bright \textit{et al} 2018, their Table 1).  These previous fits have the form 
\beq
F_X(t) = F_0 \left[ \frac{t + t_0}{t_X}\right]^{-5/3} ~~~ \text{(Power-law fit)} 
\eeq
and 
\beq
F_X(t) = F_0 \exp\left[ - \left(\frac{t+t_0}{t_X}\right)\right] ~~~ \text{(Exponential fit)}.
\eeq
In these models, the fitting parameters $F_0, t_0$ and $t_X$ are unconstrained. 
Following Bright \textit{et al.}  we use the reduced chi-squared as a measure of goodness of fit: 
\beq\label{chisq}
\bar \chi^2 \equiv \frac{1}{N - d} \sum\limits^N_{i=1} \frac{\left(O_i - M_i\right)^2}{\delta O_i^2} ,
\eeq
where the $N$ observational quantities, denoted $O_i$, have associated error $\delta O_i$.  A model with $d$ free parameters produces $N$ corresponding model predictions, denoted $M_i$. {The X-ray light curve of \Asa\ has $N = 95$ data points, while $d = 2$ for both the power-law and exponential models,  $d = 4$ for our disc model. }
{The reduced chi-squared of the three best-fitting models are}
\begin{align}
&\bar \chi^2_{\rm power-law} = 1301.1/93 = 13.99, \\
&\bar \chi^2_{\rm exponential} = 794.2/93 =  8.54, \\
&\bar \chi^2_{\rm disc} = 562.8/91 = 6.19. 
\end{align}
Given the wide parameter space available to the two fitted functions, it is noteworthy that our disc model provides a significantly better overall fit to the observed X-ray data than either of the previous fitting functions {(Fig. \ref{xray})}.  
Furthermore, of the two fitted functions the $5/3$ power law is the more physically motivated (a $5/3$ power law is a direct prediction of the Rees 1988 model). When compared to this pure power-law fit the current disc model gives a strikingly improved fit.   In addition, the disc model provides the best fit at the late times, $t > 600$ days, where the three model predictions begin to strongly diverge.    Further late time X-ray observations of ASASSN-14li would be of great interest, allowing a yet more precise testing of the three models. 

The qualitative properties of the evolving disc X-ray flux are nicely predicted by equation (\ref{FX3}).  After an initial phase of rising X-ray flux (whose properties are set entirely by the choice of initial disc conditions), there is  a pronounced light curve decay.  As may be clearly seen in {Fig.\  (\ref{xray}),} the rate at which the disc X-ray flux falls off is faster than the 5/3 power law, but  slower than the pure exponential exponential decay.   This is in accord with an evolving flux with a time dependence given by equation (\ref{FX3}).  In \S \ref{analytical}, we give a full quantitative comparison of the observed X-ray light curves with the analytical results of \S\ref{analytic}.  

Although the fit is good, the model still produces a relatively large formal reduced chi-squared statistic.  This is due to the intrinsically short timescale variations of the X-ray flux, which cannot be accounted for in a simple decay model.  {These short time-scale fluctuations may be intrinsic to accretion, produced in a jet, or caused by matter returning to the disc from large radius orbits after the disruption (fallback).}   It is these short timescale fluctuations that are the dominant source of model discrepancy, so it is not possible to significantly improve on this reduced chi-squared statistic by further adjustment of other parameters in a smooth disc model.

\subsection{UV light curve}\label{fuvsec}

\begin{figure*}
 \centering
  \includegraphics[width=.8\textwidth]{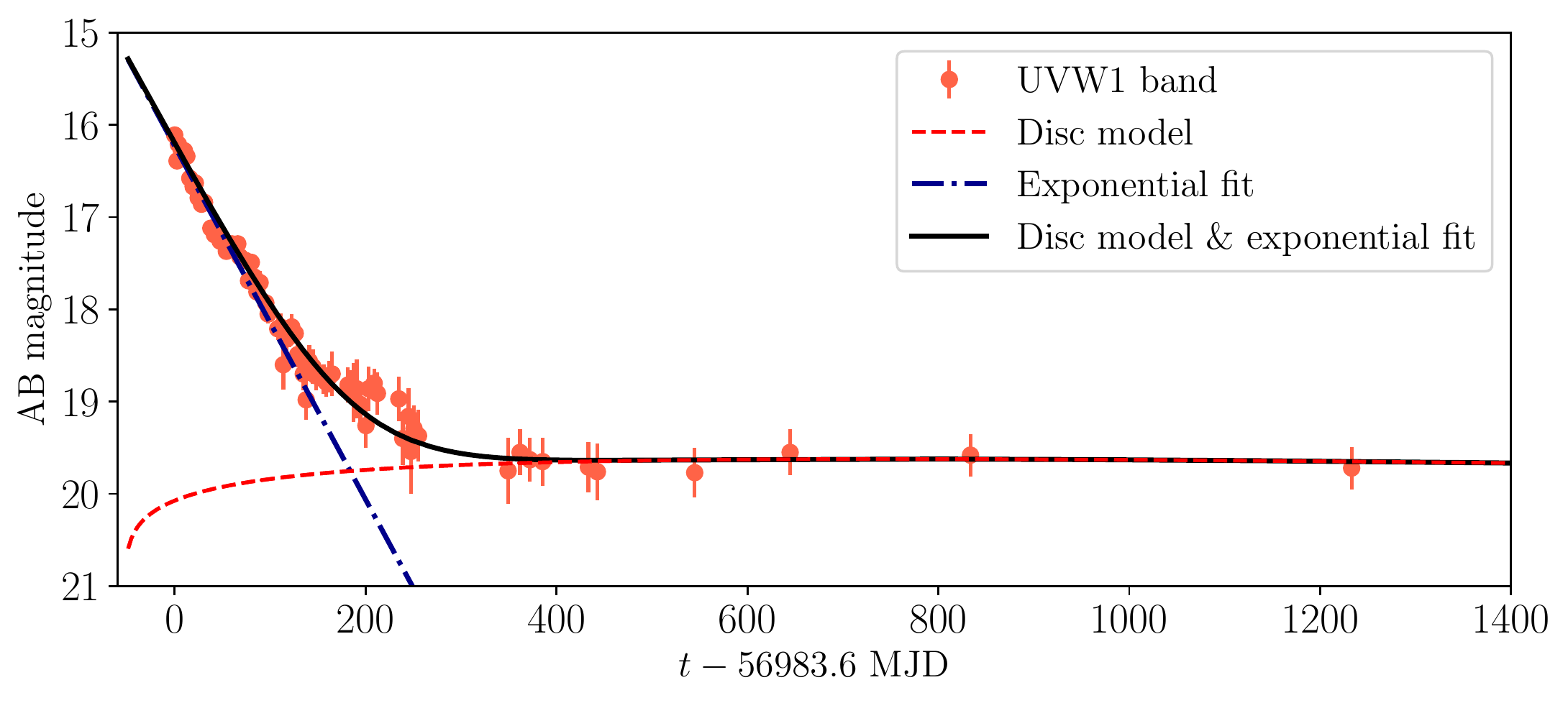} 
 \caption{Fiducial disc model fit (dashed red curve) to the evolving UV light curve of \Asa\ in the UVW1 band (Brown \textit{et al} 2017, Van Velzen \textit{et al}. 2019). The early time $t \lesssim 200$ days UV emission is not well fit by pure disc emission, but requires a secondary, exponentially decaying, component (blue dot dashed curve, see text). The disc and exponentially decaying components are assumed to sum independently. At times $t \gtrsim 200$ days the total UV emission (black solid curve) is completely dominated by disc emission, the properties of which are consistent with the analytical reasoning of section \ref{spectralshape}. }
 \label{fuv}
\end{figure*}

{Figure (\ref{fuv}) } is a plot of the evolving flux of ASASSN-14li in the UVW1 band (Brown \textit{et al}. 2017, Van Velzen \textit{et al}. 2019). The UVW1 band has a central wavelength of $\bar \lambda = 260$ nm.  The light curve is expressed as an evolving AB magnitude.   
The best fit UVW1 light curve, calculated with equation (\ref{magdef}), is plotted as a red dashed curve.   
Unlike the evolving X-ray luminosity, the observed UV emission of ASASSN-14li cannot be fit by pure disc emission at the earliest observed times.   At early times we require a second, ad hoc, rapidly-decaying, component. To the first 100 days of observations, we fit an exponentially decaying component (blue dot-dashed curve) taken from Holoein \textit{et al}. (2016):
\beq
F_\nu^\text{exp} = F_0 \exp\left[-(t-t_D)/t_{UV}\right].
\eeq 
We find a best fit of $t_{UV} = 54$ days.   Assuming that the flux from these two components sum linearly, the total observed flux of the TDE is given by 
\beq
F^{\text{total}}_\nu = F_\nu^\text{disc} + F_\nu^\text{exp}.  
\eeq
This is plotted as the ``total light curve'' (black solid curve) in Fig.\ (\ref{fuv}).   The fit is excellent.   The dominance of an UV component during the early evolutionary stages ($t < 200$ days {post observation}), not associated with thermal disc emission, has been noted previously (Pasham {\it et al.} 2017, Van Velzen {\it et al.} 2019).   Our results  are in accord with these findings. 

After an initial phase dominated by this non-disc component, the evolving UV flux is strongly disc dominated.    The final 1000 days of ASASSN-14li UV flux are almost completely unchanging with time.   As argued in \S\ref{spectralshape}, this is precisely what would be expected for bandpasses at frequencies below the thermal spectral peak, a condition satisfied here (see {Fig. \ref{examplespectrum}). }

\subsection{Characteristic observed temperatures}
In fitting observations of tidal disruption events taken in a single band, modellers often use a single characteristic temperature.  For example, two distinct characteristic UV and X-ray temperatures of ASSASN-14li were reported in Brown \textit{et al}. (2017).     A single temperature is generally inferred by fitting a pure blackbody function to the observed spectrum i.e., a Planck function (\ref{planck}),  neglecting photon red-shifting ($f \simeq 1$) and assuming a constant temperature $T$. 
{Figure (\ref{singlecolour})}  demonstrates that when observed with a single band pass, a {\em composite} disc spectrum of the form used in this work is practically indistinguishable from the spectrum of a single-temperature model.  This also shows how our composite disc spectrum is consistent with both the X-ray and UV inferred temperatures of ASASSN-14li  (Brown \textit{et al}. 2017).     

{If one fits a sequence of standard Planckian profiles to an evolving narrow band TDE light curve, two time dependent quantities can be calculated: an evolving single temperature and  emission radius.  The latter is,  in effect, an evolving overall normalisation.  A time-dependent  UV light curve  (e.g, one with  slowly decaying magnitude, $F_\nu \sim t^{-4n/7}$, eq. \ref{flatsection}) would then be interpreted by this procedure as a uniform temperature blackbody emission with a slowly shrinking radius.   Very recently, a TDE-candidate ASASSN-18jd (Neustadt {\it et al}. 2019) was observed showing exactly this feature, and was analysed by the technique we have described.    In common with \Asa, ASASSN-18jd was discovered after a $\sim 120$ day seasonal gap in observations, and appears to have already entered its UV plateau phase.   The shallow decay of the UV light curve was interpreted by Neustadt {\it et al.} as a constant temperature blackbody which is slowly {\em contracting.}    We suggest here a second possible interpretation: that ASASSN-18jd may be another example of a TDE exhibiting an extended disc-dominated phase, and that this disc is cooling and {\em expanding.} }

{Furthermore, van Velzen {\it et al} (2019) surveyed a population of 8 TDEs detected at optical and UV wavelengths. They found evidence for a late-time UV plateau in the 6 best-observed TDEs in their sample, suggesting that the majority of TDEs  undergo a disc-dominated phase.     }
\begin{figure}
 \centering
  \includegraphics[width=.5\textwidth]{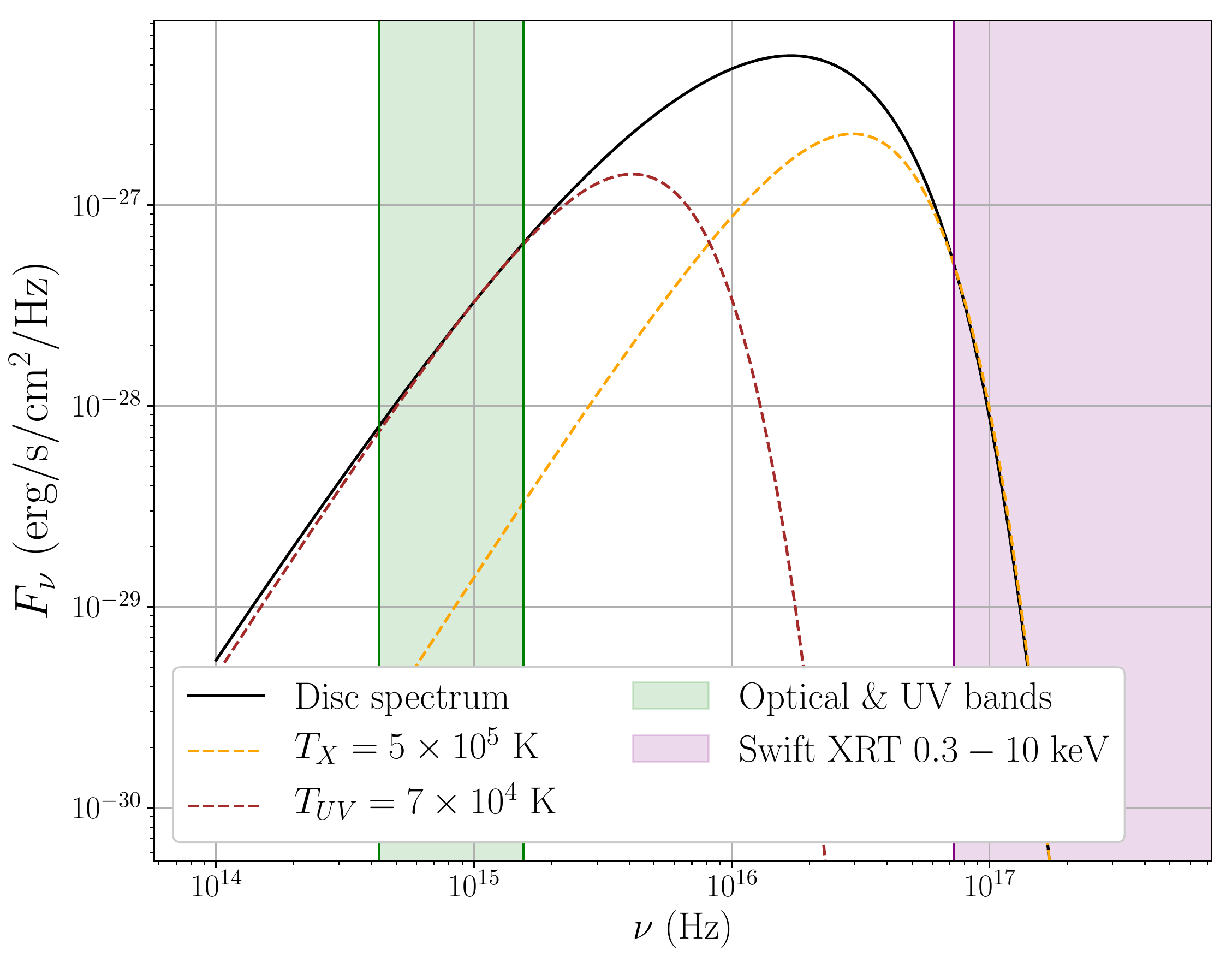} 
 \caption{A comparison between a composite disc spectrum (black solid curve, Fig. \ref{examplespectrum}) and two single colour blackbody spectrums (brown and yellow dashed curves), with `characteristic'  temperatures denoted on  plot.  When observed through either optical or X-ray bands a composite spectrum of the sort studied in this paper is largely indistinguishable from that of a single colour spectrum. }
 \label{singlecolour}
\end{figure}

\subsection{Model limitations: a summary}\label{bol}

{In common with all analytic models of TDE evolution, our model has necessarily been simplified in a number of ways.   While we have included general relativistic effects, we have neglected radiative transfer in the disc atmosphere.   While these effects are likely to be small,  it is of course possible that our best fit model parameters would be affected by their inclusion.   The general properties of relativistic thin disc light curves are likely to be insensitive to this level of detail:  the UV plateau results from the spreading and cooling of the disc on large scales, whilst the X-ray light curve is described by equation (\ref{FX3}).   The peak effective temperature $\T_p$ could be somewhat modified to include a colour-correction for Compton scattering, but the functional form itself will be unaltered: the explicit time dependence enters through the solution of the underlying disc equations. 

Further potential sources of spectral modification could in principle arise from any irradiation of the disc by an external source (e.g, a jet), or even self-irradiation by disc photons which have been gravitationally deflected to such a degree that they illuminate the disc.   (This final effect becomes important if the angular momentum parameter of the black hole approaches unity.)   Finally, whilst it appears that the early time non-disc emission is restricted to UV wavelengths,  it is not possible to rule out the possibility that there could be a non-thermal component at X-ray frequencies at the earliest times. 

\begin{figure}
 \centering
  \includegraphics[width=.5\textwidth]{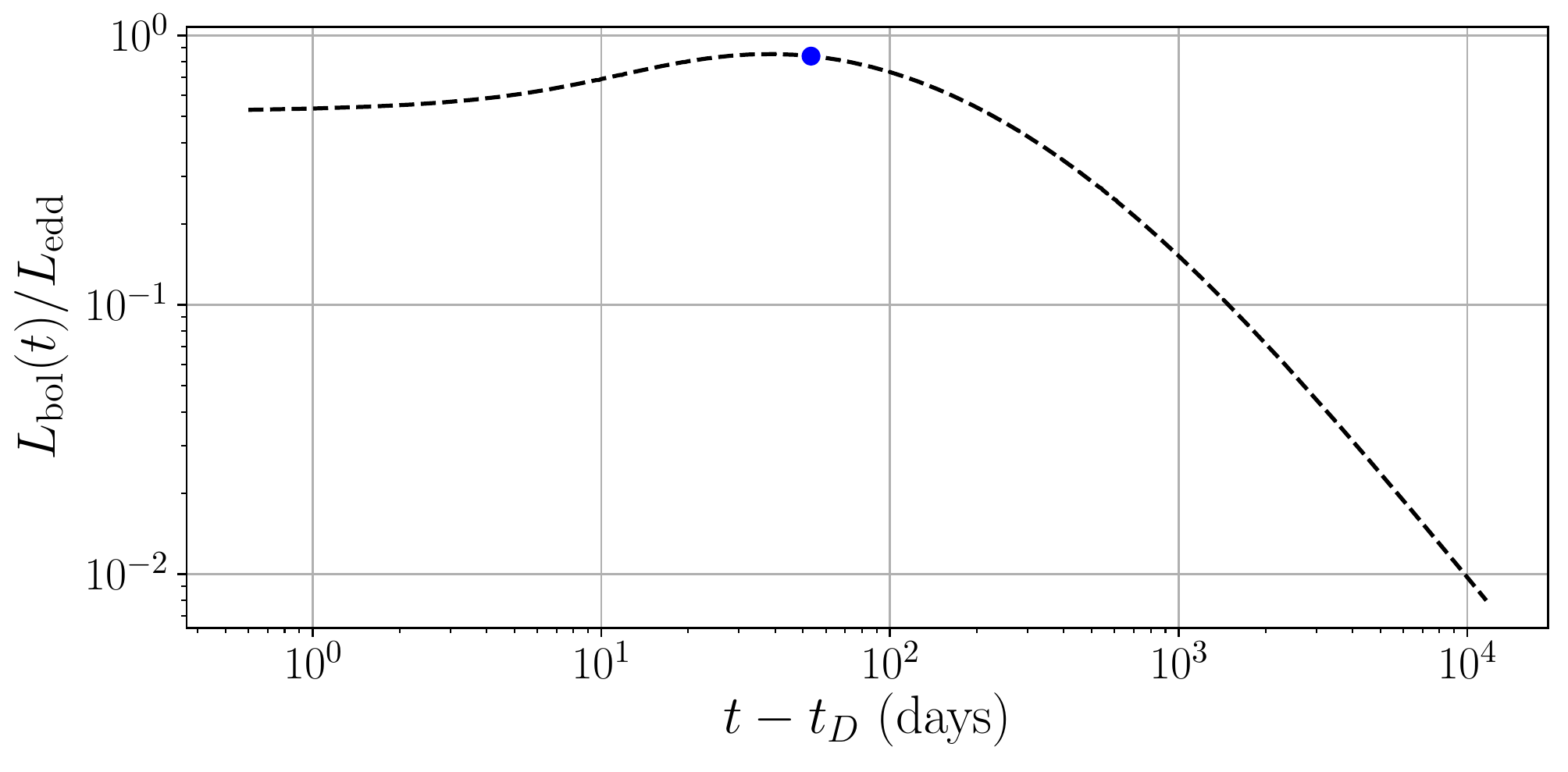} 
 \caption{The {computed} bolometric luminosity of our best fit disc model, plotted in units of the Eddington luminosity for our best fit black hole mass. The blue dot coincides with the time of the first recorded observation of \Asa . The peak of this light curve corresponds to a bolometric luminosity of $L_{\rm peak} \simeq 0.85\, L_{\rm Edd}$. }
 \label{bolometric}
\end{figure}

A fundamental assumption of our model is that the disc is thin, and can therefore be described by the relativistic evolution equation (\ref{fund}).  This assumption will breakdown if the ratio of the bolometric luminosity to the Eddington luminosity significantly exceeds unity.  This may be   checked numerically.   The bolometric luminosity is found by integrating the locally emitted flux over the disc surface
\beq
L_{\rm bol}(t) = 2\pi \int_{r_I}^{\infty} ({g_{rr} g_{\phi\phi}})^{1/2} \gamma^\phi(r,a) \, 2 \sigma \widetilde T^4(r,t) \, {\rm d}r . 
\eeq
For our best-fit black hole mass, $M =  1.85\times 10^6 M_\odot$, the Eddington luminosity is
\beq
L_{\rm Edd} = 2.33 \times 10^{44} \, {\rm erg/s} \simeq 5.9 \times 10^{10} L_{\odot}.
\eeq
The bolometric light curve {of our best fit disc model} is shown in Figure (\ref{bolometric}).  Whilst at the earliest times  the disc luminosity approaches the Eddington limit, peaking at $L_{\rm peak}/L_{\rm Edd} \simeq 0.85$ (the blue dot corresponds with the time of the first recorded \Asa\ observation), the disc is  found to be sub-Eddington thereafter.  {With a peak luminosity of $L_{\rm peak} = 0.85 L_{\rm Edd}$, we note that corrections to the thin disc model need not be negligible, as the ``slim disc'' regime is entered (Abramowicz et al. 1988).   However, it is unlikely that such modifications will result in transformative changes in the observed light curves, as opposed to slightly altering our thin disc model fits.   Along with the effects of radiative transfer and radiatively coupled outflows, slim disc modelling represents a promising avenue for refining the present treatment. }

As was earlier suggested, both the small disc mass and the early time UV flux could result from a super-Eddington outflow, initiated at times predating the first observation.  If so, it is not a surprise that the disc luminosity at the earliest times is {close to the Eddington luminosity}.   If representative of the wider TDE population, our results suggest that a relatively small fraction of the stellar debris eventually settles into a disc at radii close to the central black hole.  This will be particularly true of {discs formed from} TDEs around lower mass black holes, which have both a lower Eddington luminosities and larger bolometric luminosities for a given disc mass (note that the scaling relations for time dependent discs differ from classical steady state models, as in the former $\dot M$ is a dynamical variable). }

\begin{figure*}
 \centering
  \includegraphics[width=.8\textwidth]{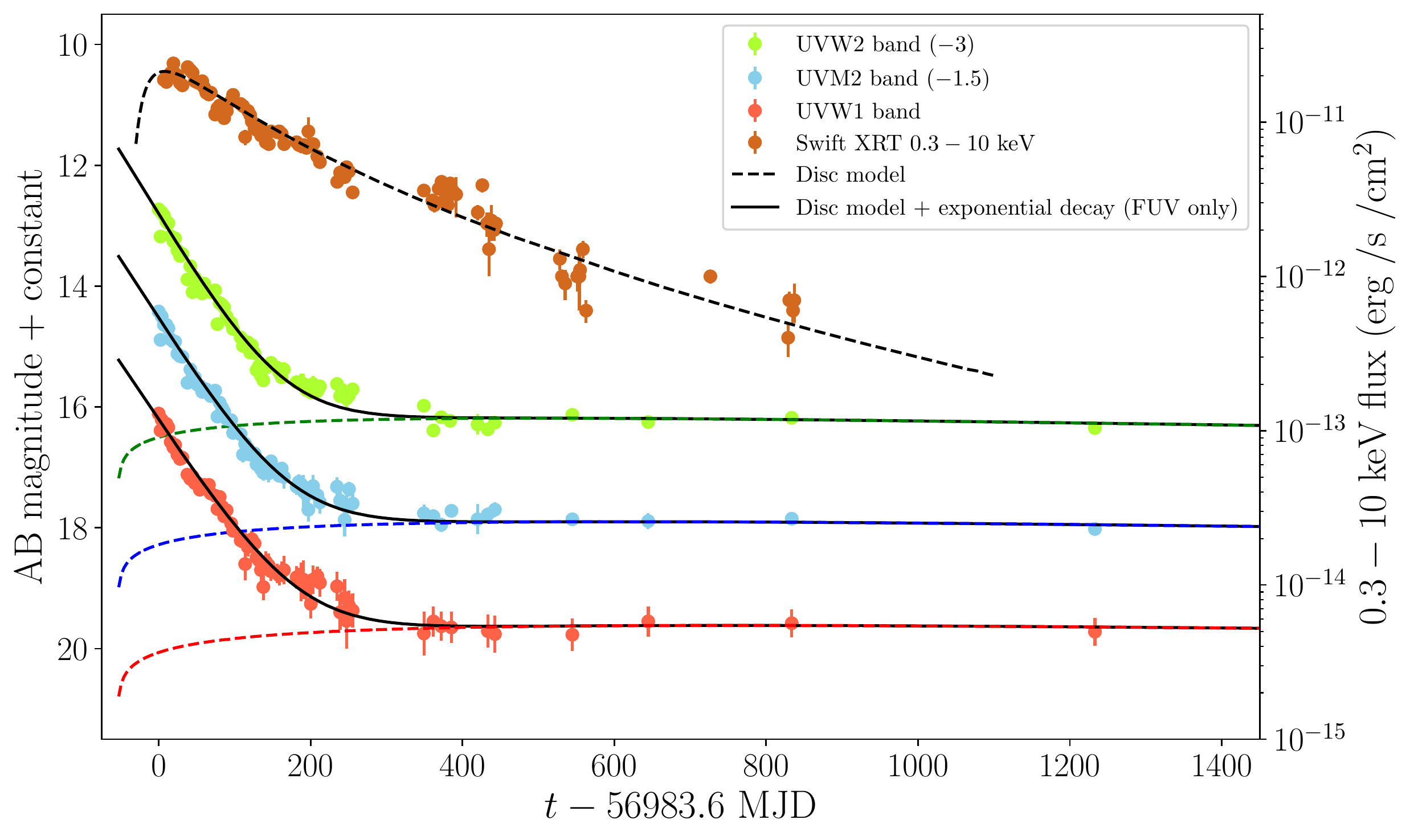} 
 \caption{Comparison of the best fit disc light curves with the observed ASASSN-14li light curves at both X-ray (right scale) and UV (left scale) frequencies (UV data: Brown {\it et al}. 2017, Van Velzen {\it et al}. 2019; X-ray data: Bright {\it et al}. 2018). The disc model light curves are shown by the dashed curves, at UV wavelengths there is an additional exponentially declining component at early times. All four disc light curves result from the same disc model with only four fitted parameters.}
 \label{total}
\end{figure*}

\subsection{Summary of numerical model}

In this section, we have determined  best fit disc light curves at several different wavelength bands for a relativistic thin disc evolving with a finite ISCO stress.  
These model light curves are compared to the observed light curves of the tidal disruption event ASASSN-14li.  By solving the fully relativistic thin-disc equation (\ref{fund}) and performing ray tracing calculations, this model includes strong gravity effects.   Figure (\ref{total}) shows the evolving emission in 4 different observational bands, three at UV wavelengths and one corresponding to the \textit{Swift} X-ray telescope 0.3 -- 10 keV band. The dashed curves represent the contributions of the evolving disc to the emitted flux. The solid lines represent a combined disc and exponentially decaying component, only relevant at UV wavelengths. 

We have found a number of interesting results. The evolving  X-ray flux is well described at all times by an evolving thin disc  with a finite ISCO stress.   The model requires no other spectral components and gives a significantly better fit to the observations than either the canonical Rees power law model or an exponentially decaying model.  The properties of this X-ray light curve are in good qualitative agreement with the analytical results of \S\ref{analytic}: the X-ray flux of the disc model decays at a faster rate than that of a pure power law, but at a slower rate than a pure exponential decay.  

The UV emission of ASASSN-14li at the earliest times appears not to be the result of thermal emission from a turbulently evolving thin disc.   Instead,  it is dominated by a much more rapidly decaying, secondary component.  After $\sim 200$ days however, the ASASSN-14li UV flux undergoes a 1000 day plateau, and is extremely well described by thin disc emission. 

This means that for approximately 1000 days of observations, the entirety of the ASASSN-14li emission at both UV and X-ray wavelengths can be successfully reproduced by a relativistic thin disc model with nine total parameters, of which only four are inferred from the observations. Our results demonstrate that while the behaviour of a TDE at its earliest times may be complex, the long term observed properties of a TDE can be very well described by a simple relativistic thin disc with a finite ISCO stress.  This suggests the exciting possibility that future modelling of well observed TDEs will be able to constrain the properties of a population of otherwise quiescent super massive black holes.

\section{Properties of the central black hole}\label{BHproperties}

\begin{table}
\renewcommand{\arraystretch}{2}
\centering
\begin{tabular}{|p{1.3cm}|p{2.1cm}|p{1.3cm}| p{1.3 cm} |}
\hline
$a/r_g$ & $M/M_{\sun}$  &  $\bar\chi_{\text{min}}^2$\\ \hline\hline
$0.9$ & $2.00^{+0.05}_{-0.1}\times10^6$ &  6.57 \\ \hline 
$0.75$ & $1.93^{+0.03}_{-0.04} \times 10^6$ & $6.15$ \\ \hline 
$0.5$ & $1.85^{+0.03}_{-0.04} \times 10^6$   & $6.19$  \\ \hline
$ 0 $ & $1.66^{+0.06}_{-0.06} \times 10^{6}$  & $6.40$ \\ \hline
$-0.5$ & $1.62^{+0.08}_{-0.07} \times 10^6$  & $6.49$ \\ \hline
$-0.9$ & $1.55^{+0.1}_{-0.1} \times 10^6$   & $6.61$  \\ \hline 
\end{tabular}
\caption{Best fit black hole masses $M$ and the minimum X-ray light curve reduced chi-squared  $\bar\chi_{\text{min}}^2$ for different black hole spin parameters. The non-fitted parameters of Table \ref{table2} are unchanged. }
\label{tablespin}
\end{table}

To analyse the effects of varying the black hole spin on the best fit ASASSN-14li parameters, we investigate the disc evolution equation (\ref{fund}), along with ray-tracing calculations (\ref{redshift}), for a variety of black hole spins. For each black hole spin we refit the 4 parameters $t_v, \, t_D,\, M_d$ and $M$ to the X-ray light curve and late time UV emission. Of these four parameters, the viscous timescale $t_v$, disc mass $M_d$ and time offset $t_D$ are the most poorly constrained by theory or observations.  For each black hole spin, we present in table \ref{tablespin} the best fit black hole mass $M$, and the minimum reduced chi-squared $\bar \chi_\text{min} ^2$ of the fitted X-ray light curve (equation \ref{chisq}).  

The value of the minimum reduced chi-squared of the fitted X-ray light curves is relatively insensitive to the black hole spin (Table \ref{tablespin}). This is due to the fact that residual short-timescale fluctuations in the ASASSN-14li X-ray light curve dominate the model-data discrepancy. 
This is further compounded by an intrinsic parameter degeneracy between the black holes mass and spin.    The peak temperature of a finite ISCO stress disc, which is the physically relevant temperature for the X-ray flux (eq.\  \ref{FX3}), depends strongly on the physical location of the ISCO.   Increasing the (prograde) rotation of the black hole reduces the ISCO location in units of $r_g$, whereas increasing the black hole mass increases the magnitude of $r_g$ in physical units.  This is reflected in the positive correlation of the best fit black hole masses and spins  (Table \ref{tablespin}).   {With these caveats, a best-fit black hole spin lies in the range $0.5 < a/r_g < 0.75$.   This is much more uncertain than other fitted parameters}. 

We can use the physically allowed spin range ($|a/r_g| < 1$) to calculate bounds on the black hole mass. An important result of our analysis is that the central black hole mass of ASASSN-14li can be tightly constrained. Over the dimensionless spin range $(-0.9 < a/r_g < 0.9)$ considered here, ({we do not extend this range to $a/r_g = \pm 0.998$ as the neglected returning radiation would become significant in this limit})  the best fit black hole mass varies by only a factor of 1.3, varying from $1.55  \times 10^6 M_{\sun}$ for rapid retrograde rotation  $a/r_g = -0.9$, to $2 \times 10^6 M_{\sun}$ for rapid prograde rotation $a/r_g = 0.9$.    This best fit mass range lies at the lower end of the range inferred from galactic bulge mass measurements ($10^{6.7 \pm 0.6} M_{\sun}$, Mendel \textit{et al}. 2014), {but is in agreement with the mass inferred from velocity dispersion measurements ($M = 10^{6.23 \pm 0.4} M_{\odot}$, Wevers {\it et al.} 2017). }

Previous estimates based on the observed ASASSN-14li light curves are much more in keeping with our findings, with Miller \textit{et al}. (2015) inferring a black hole mass of $M \sim 2 \times 10^6 M_{\sun}$ based on Eddington limit arguments and the observed X-ray luminosity. Finally, our inferred mass range is also consistent with the QPO bound published by Pasham \textit{et al}. (2019),  $M \lesssim 2 \times 10^6 M_{\sun}$. It therefore  appears that very distinct modelling techniques are converging onto a single black hole mass for the ASASSN-14li system, of order $M = 2 \times 10^6 M_{\sun}$.

\section{Finite versus Vanishing ISCO stress}\label{FvsVstress}

\subsection{High photon energy expansion}\label{VSexpansion}
If the stress vanishes at the ISCO, so too does the local disc temperature.  This means that the hottest disc regions -- those which contribute to the high energy Wien tail --  take the geometric form of a ring at a disc radius exterior to the ISCO, within the body of the disc.   To calculate the high frequency spectrum in this case, we follow \S\ref{QWlim} and expand the effective temperature about its maximum, assuming a face-on disc orientation.   Now, the maximum temperature is now  longer at the integration limit, and the expansion takes the form 
\beq
{{1}\over{k_B \widetilde T}} = {{1}\over{k_B \widetilde T_p}}  + {{1}\over{2}} {{\partial^2}\over{\partial R^2}} \left[ {{1}\over{k_B \widetilde T}} \right]_{R_p} (R - R_p)^2 + ... 
\eeq
We next define a dimensionless normalisation parameter 
\beq
\xi \equiv R_p^2 k_B \widetilde T_p  {{\partial^2}\over{\partial R^2}} \left[ {{1}\over{k_B \widetilde T}} \right]_{R_p} .
\eeq
The leading order flux integral is then
\begin{multline}
F_\nu  =  {\frac{4\pi h\nu_o^3}{D^2 c^2}} \exp\left(- {{h\nu_o}\over{k_B \widetilde T_p}} \right) \times 
 \\ \int_{-\infty}^{\infty} \, R  \exp\left[-  {{\xi h\nu_o}\over{2 k_B \widetilde T_p}} {{(R-R_p)^2}\over{R_p^2}}\right]   \, \text{d} R .
\end{multline}
This integral may be evaluated to give
\beq
F_\nu =  {\frac{4\pi h\nu_o^3}{ c^2}}\sqrt{\frac{2\pi}{\xi}} \left({{R_p}\over{D}}\right)^2 \left({{k_B \widetilde T_p}\over{ h\nu_o}}\right)^{1/2} \exp\left(- {{h\nu_o}\over{k_B \widetilde T_p}} \right) . 
\eeq
The X-ray luminosity is obtained by integrating this expression over the observer frequency band of interest.    To leading order this is
\beq
F_X =  {\frac{4\pi h\nu_l^4}{ c^2}}\sqrt{\frac{2\pi}{\xi}} \left({{R_p}\over{D}}\right)^2 \left({{k_B \widetilde T_p}\over{ h\nu_l}}\right)^{3/2} \exp\left(- {{h\nu_l}\over{k_B \widetilde T_p}} \right) .
\eeq
As in \S \ref{QWlim} we assume a temperature time dependence of the form $\widetilde T_p \propto \tau^{-n/4}$.      A disc with a vanishing ISCO stress therefore has an evolving X-ray light curve that takes the form
\beq
F_X(t) = F_0 \left({{t+t_0}\over{t_X}}\right)^{-3n/8} \exp\left[ -\left({{t+t_0}\over{t_X}}\right)^{n/4} \right].
\eeq
where $t_0$ and $t_X$ are constants.   We remind the reader that for a vanishing ISCO stress,  $n > 1$. 

\begin{figure}
  \includegraphics[width=.5\textwidth]{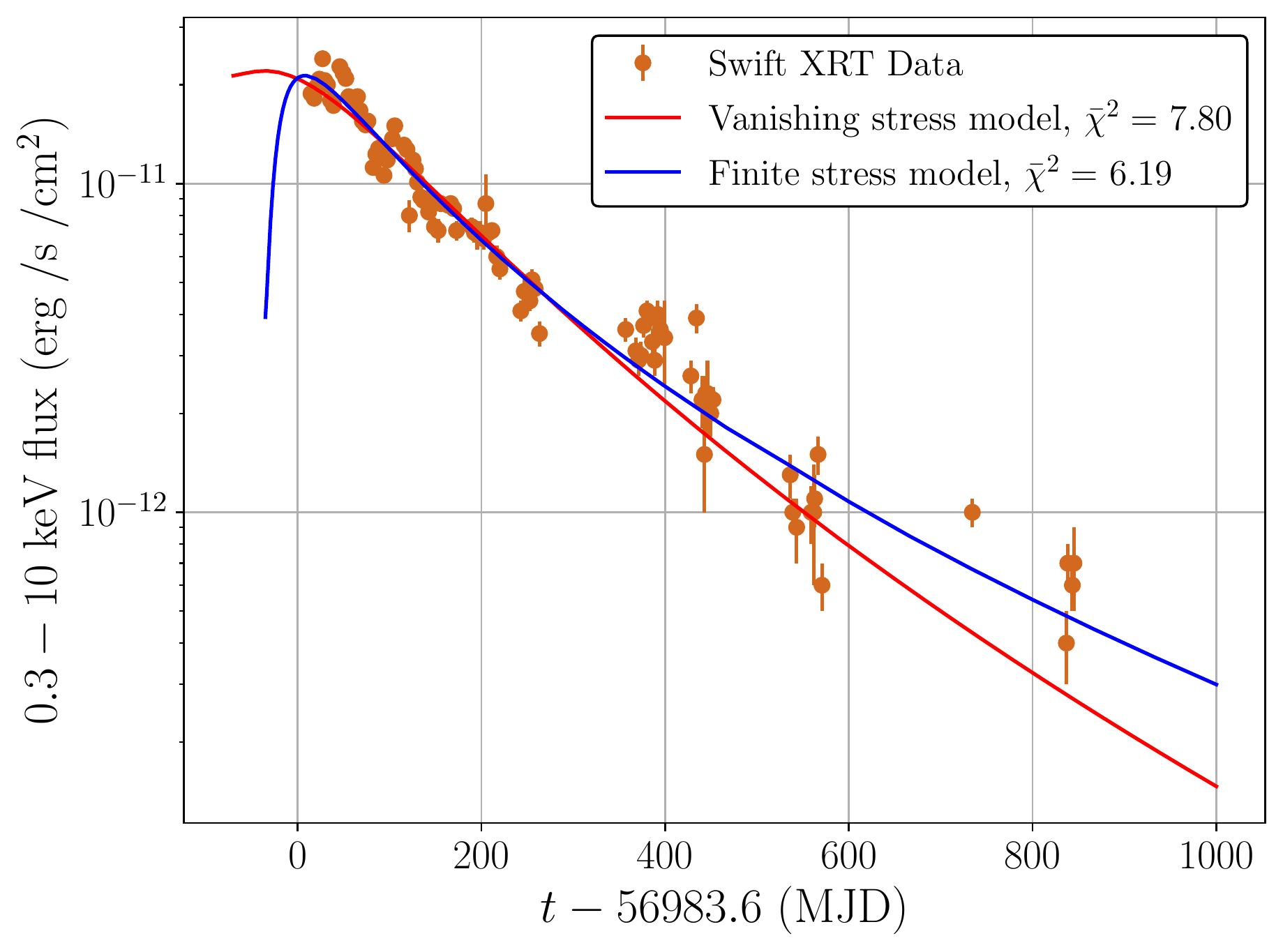} 
 \caption{ A comparison of the best fit X-ray light curves from discs evolving with  finite (blue) and vanishing (red) ISCO stresses, with the observed X-ray light curve of ASASSN-14li (Bright {\it et al}. 2018). The steeper fall-off of the vanishing ISCO stress disc solution can be understood by its larger decay index $n$ (see text). A disc with a finite ISCO stress produces a better fit to the observed \Asa\ light curve.  }
 \label{FSVScompare}
\end{figure}

\subsection{Best fit light curves}
Figure (\ref{FSVScompare}) shows the best fit X-ray light curve of a vanishing ISCO stress disc.  This light curve was found with the same non-fitted parameters of \S \ref{numerical}  (Table \ref{table2}), except for the ISCO stress parameter $\gamma$, which  is in effect infinite for a vanishing ISCO stress (MB2).  Next, the parameters of Table (\ref{table1}) were simultaneously fit to the ASASSN-14li X-ray and UV light curves. 

{The best fit parameters for the vanishing ISCO stress disc are: $M = 1.8 \pm 0.1 \times10^6 M_{\odot}$, $M_d = 0.30 \pm 0.04 M_{\odot}$, $t_v = 400 \pm 20$ days and $t_D = 110 \pm 10$ days. The best-fit values for the vanishing ISCO stress disc are less certain than the finite ISCO stress disc as a result of the overall poorer fit, $\bar\chi^2_{\rm vanishing \ stress} = 709.8/91$.  }
As can be seen in Figure (\ref{FSVScompare}), the best fit vanishing and finite ISCO stress light curves are almost indistinguishable for the first $\sim 300$ days of observations, deviating only at later times. This late time deviation is entirely a result of the different luminosity power law exponents associated with vanishing and finite ISCO stress discs.  With a power law exponent  of $n \approx 1.2$, the late time vanishing ISCO stress X-ray light curve becomes dominated by a $F_X \sim t^{-0.45} \exp\left( - t^{0.3} \right)$ time dependence. The correspondingly shallower decline of a finite ISCO stress X-ray light curve results from its $n \approx 0.8$ exponent, leading to a $F_X \sim t^{-0.4} \exp  \left(-t^{0.2}\right)$ late time profile.  It is the different exponents within the exponentials which are responsible for the late time divergence of the two light curves.

Since ASASSN-14li has been observed for $\sim 900$ days, there are observations at sufficiently late times to differentiate between the two models.    The X-ray light curve of ASASSN-14li is clearly better fit with a disc evolving with a finite ISCO stress than a disc with a vanishing ISCO stress (Fig. \ref{FSVScompare}), a distinction that can only be seen at times $t > 400$ days, where the two models cleanly diverge. {An additional problem with the vanishing ISCO stress disc model is that to produce an acceptable X-ray flux, the disc must be super-Eddington.  The vanishing ISCO stress discs bolometric luminosity peaks at $L_{\rm peak} = 1.31 \,L_{\rm Edd}$, and $L/L_{\rm Edd} > 1$ for the first $\sim 250$ days.}   Interestingly, late time observations of evolving TDEs may be the best way to determine the nature of the ISCO boundary condition.

\section{Comparison of numerical and analytical calculations}\label{analytical}
\begin{figure}
 \centering
  \includegraphics[width=.5\textwidth]{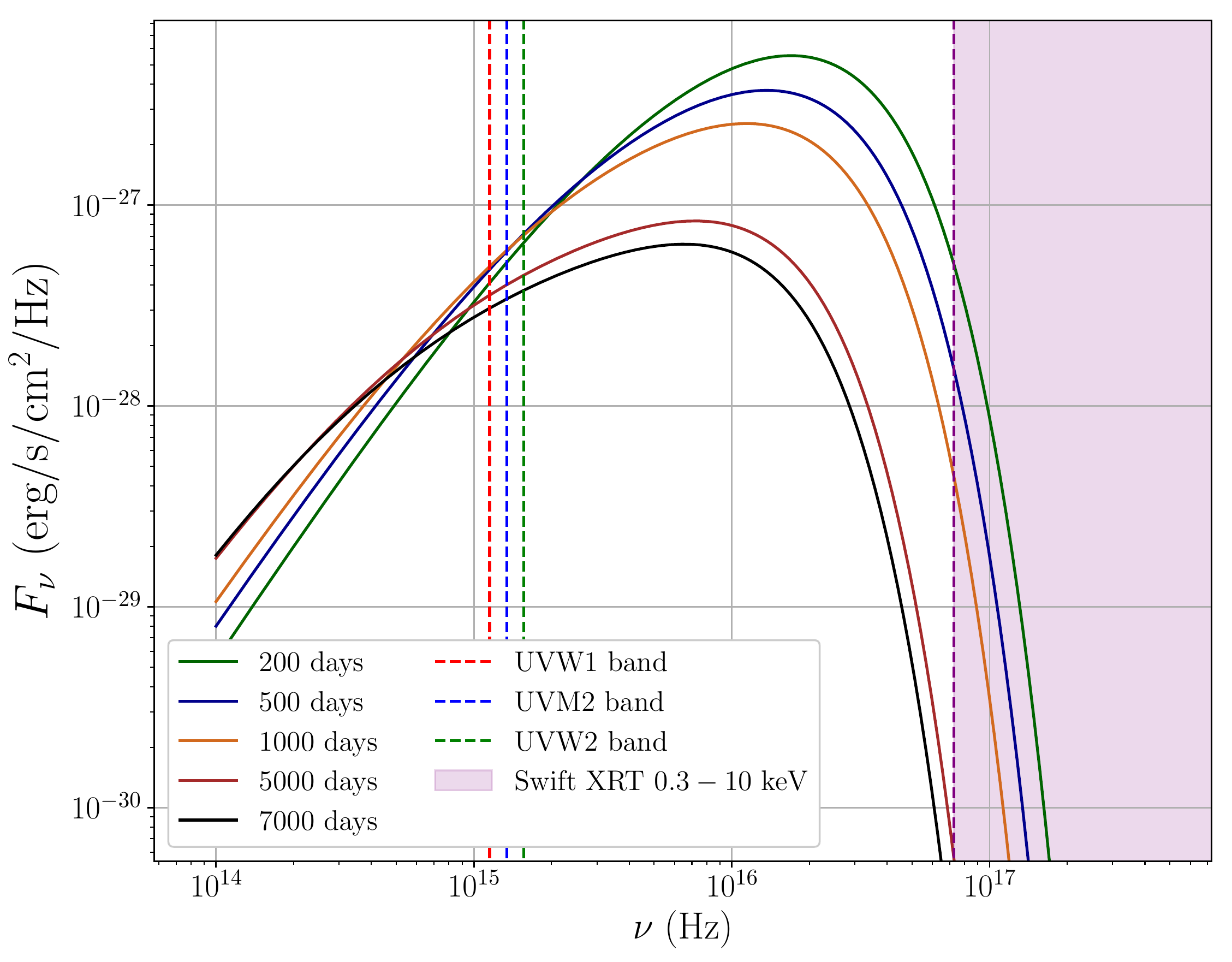} 
 \caption{The disc continuum spectrum, calculated from equation (\ref{flux}), shown at several different times (denoted on plot).  The simultaneous properties of both the (very shallow) UV and (rapidly declining) X-ray disc light curves can be simply understood from the evolution of the overall shape of the disc spectrum.  }
 \label{evolvingspectrum}
\end{figure}

\begin{figure}
 \centering
  \includegraphics[width=.45\textwidth]{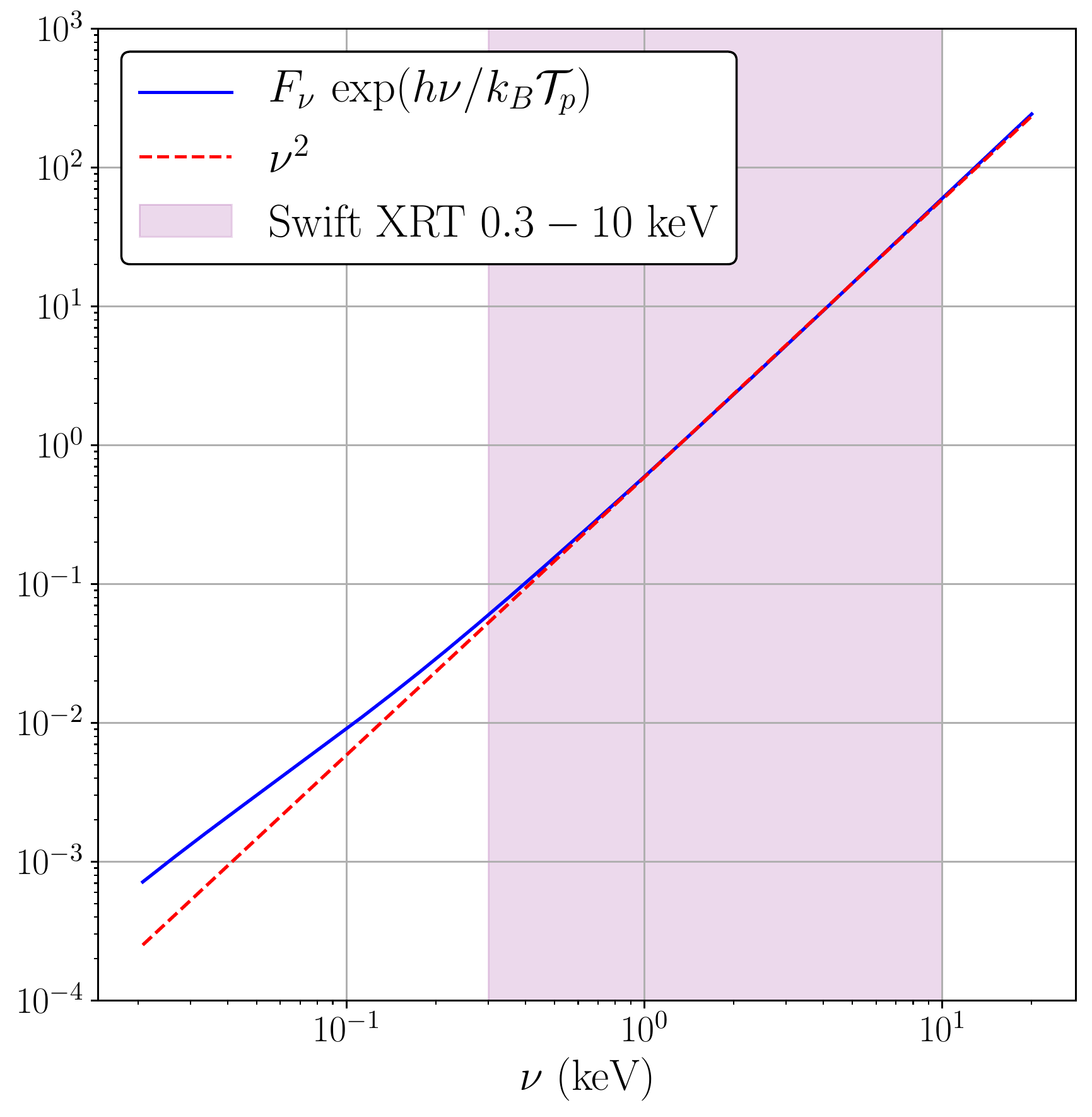} 
 \caption{Comparison of  $F_\nu \exp(h\nu/k\T_p)$ (renormalised for display) with the large $\nu$ asymptotic result for the case of a finite ISCO stress.  The solid line is from a numerical evaluation (Fig. \ref{examplespectrum}); the dotted line is the large $\nu$ analytic result (equation \ref{hespec}).}
 \label{highvspectrum}
\end{figure}

\begin{figure*}
 \centering
  \includegraphics[width=.7\textwidth]{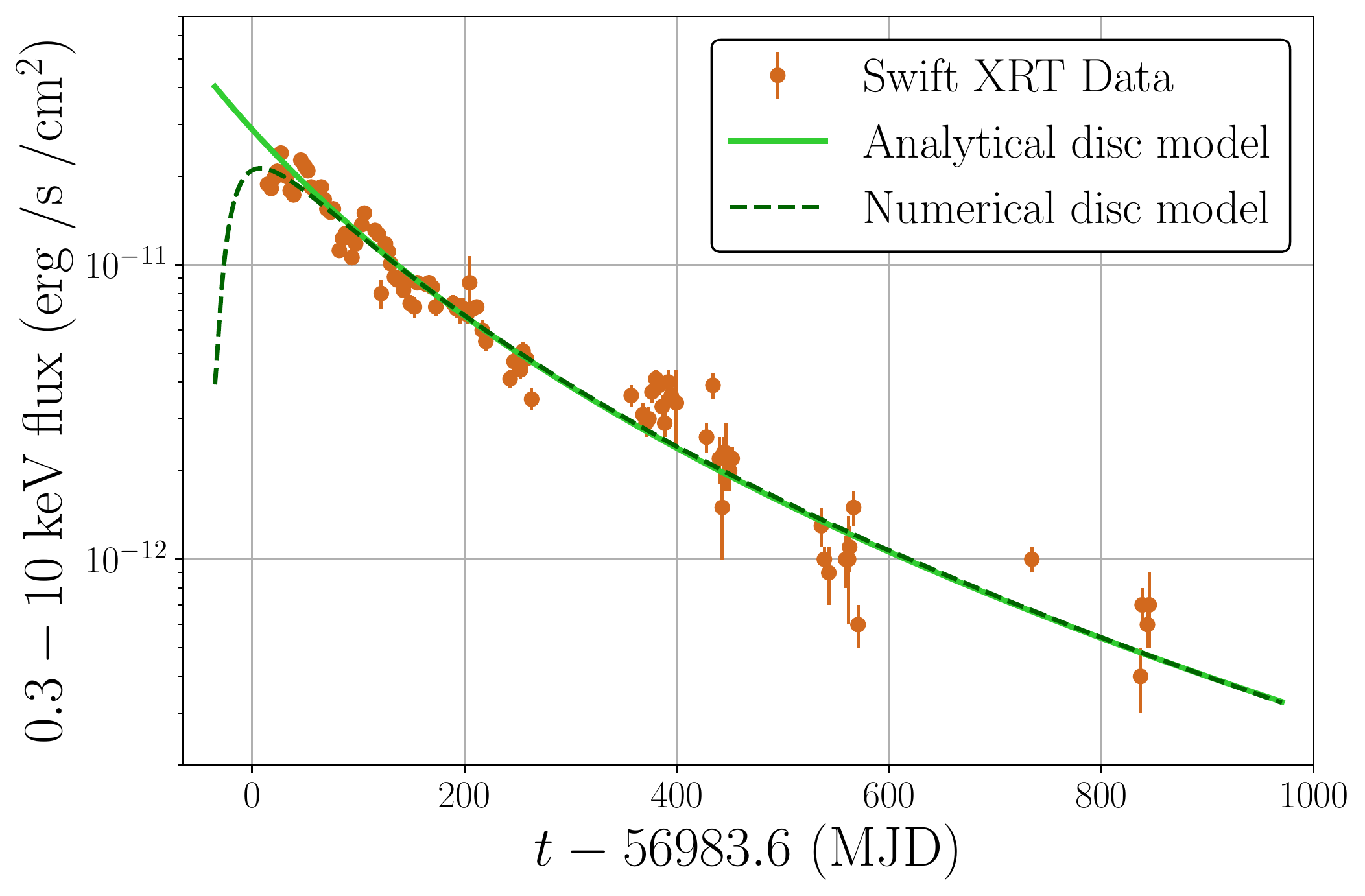} 
 \caption{Comparison between the asymptotic X-ray light curve function (eq. \ref{helum}), the best fit numerical X-ray light curve (Fig. \ref{xray}), and observed X-ray light curve of ASASSN-14li (Bright {\it et al}. 2018). The numerical X-ray light curve is extremely well described by the asymptotic function (eq. \ref{helum}). }
 \label{xrayanalytical}
\end{figure*}

We have seen that the results of the full numerical calculations of the evolving ASASSN-14li light curves are in good qualitative agreement with the analytical results of \S\ref{analytic}.   In this section we focus on a more quantitive comparison between numerical and analytical calculations. 

Not surprisingly, the behaviour of the ASASSN-14li light curves shows a strong dependence on observational band.  The UV flux undergoes a prolonged plateau, whilst the X-ray flux declines both monotonically and rapidly.   This behaviour can be understood by the relative location of the observational bands with respect to the frequency of the peak of the continuum accretion disc spectrum (figure \ref{evolvingspectrum}).  As figure (\ref{evolvingspectrum}) demonstrates, the UV bands probe frequencies smaller than peak of the disc spectrum.   At these frequencies, the dual effects of a temporally decreasing peak intensity, coupled with the decline of the peak frequency itself leads to narrow band observations in which the disc flux  plateaus for prolonged periods.    This is particularly pronounced in the UV observations of ASASSN-14li (figure \ref{fuv}). 

We emphasise that this UV light curve plateau should be a signature feature of all tidal disruption events which undergo an evolutionary  phase dominated by disc emission. Once a TDE system has settled down into an evolving thin disc (after an initial phase of UV emission dominated by some non-thermal component), physically reasonable values of the parameters $t_v, \, M_d$ and $M$ will cause the disc spectral peak to lie at `extreme' ultra-violet frequencies, which are much higher than the UV observational bands.  The late time observations ($t \gtrsim 200$ days) of a wide range of TDEs should therefore display UV light curve plateaus.  

As seen clearly in figure (\ref{evolvingspectrum}), the observed ASASSN-14li X-ray flux stems from the quasi-Wien tail of the disc spectrum.   This allows for a detailed comparison between the high-photon-energy asymptotic expansions presented in \S\ref{analytic} (equations \ref{flux5} \& \ref{FX3}), and the best fit numerical spectrum, and evolving X-ray light curve, of \S\ref{numerical} (figures \ref{examplespectrum} \& \ref{xray}).  Figure (\ref{highvspectrum}) is a comparison between the high-photon-energy spectrum of the fiducial disc model (best fit parameters from Table \ref{table1}), and the functional form of the asymptotic expansion (equation \ref{flux5}).  The observed flux has been renormalised for display purposes, but the frequency dependence is of course unaltered.   The asymptotic flux expression  
\beq\label{hespec}
F_\nu \simeq \nu^2 \exp(-h\nu/k_B\T_p),
\eeq
where $\T_p$ is the maximum effective temperature in the disc at the time of fitting, is clearly an excellent fit over the entire \textit{Swift} XRT band.  

Yet more striking is a quantitative comparison between the mathematical asymptotic expansion for the evolving X-ray light curve (eq. \ref{FX3}) and both the observed X-ray flux of ASASSN-14li, with the results of the full numerical disc model (Fig.\ [\ref{xrayanalytical}]).   It is only at the very earliest times that equation (\ref{FX3}) and the full numerical disc model deviate at all!    This is due to the assumption that the peak of the effective temperature distribution occurs at the inner disc edge, something which is manifestly not satisfied in the full numerical disc model at early times ({see Fig. \ref{tempevol}}).    Once the evolving disc reaches the ISCO the two results very quickly converge, and are then in excellent agreement for the entire 900 days of observations.  The value of the bolometric decay index was taken to be $n = 0.8$ during this fitting procedure, a value which in conformance with time-dependent disc theory for a finite ISCO stress disc (BM18, MB1, MB2).     Our results suggest that the observed X-ray light curves from disc-dominated TDE spectra should be well fit by the following function, which is a high energy asymptotic solution of the underlying disc emission integral {assuming a finite ISCO stress}:
\beq\label{helum}
F_X(t) = F_0 \left(\frac{t + t_0}{t_X}\right)^{-n/2}\, \exp\left[- \left(\frac{t + t_0}{t_X}\right)^{n/4}\right] .
\eeq 
 {with $n \simeq 0.8$}.

\section{Conclusions}

In this work, we have developed mathematical techniques, both numerical and analytical, to evaluate the emission profiles of a simple  evolving relativistic thin disc.  The {effective} temperature profiles are calculated on the basis of a recently developed 1D evolution equation for a disc in the Kerr geometry (BM18, MB1, MB2).   At the ISCO, we have allowed for either a vanishing or finite value of the stress.   

The relativistic equation has much in common with its Newtonian counterpart.  Indeed, the self-similar solutions of the Newtonian equation (of which there are two; Pringle 1991) serve as the foundation for understanding the properties of relativistic discs.  Crucially, it is the properties of the disc in the region near the ISCO which influences the global  disc behaviour at large times.    A vanishing ISCO stress picks out the classical Newtonian solution at late times, whereas a finite ISCO stress boundary condition results in the dominance of an outer disc solution which is ordinarily discarded.  

The results we have presented here are for the light curves of these solutions at observed UV and X-ray bandpasses. We have demonstrated that these light curves provide a {significantly improved} fit to the spectral data of ASASSN-14li, a particularly well observed TDE, {when compared to previous models}. The calculated light curves of discs evolving with a finite ISCO stress provide a better fit than those of a vanishing ISCO stress. The analysis here is a more stringent test than previous work (BM18) which focussed on fitting power law indices.

Our spectral modelling has been carried out in two complementary ways.  First, we have integrated the evolution equation (for the effective temperature) and evaluated the disc emission integral by direct numerical methods.   Second, we have used asymptotic expansion techniques to obtain a simple analytic form for the X-ray flux (eq.\ \ref{helum}), and determine general properties of the evolving UV flux.     A comparison between the observed light curves of  ASASSN-14li, numerical calculation, and analytic formula are all in good agreement. Over 1000 days of simultaneous UV and X-ray observations of \Asa\ are well fit with a relativistic thin disc model with a finite stress at the ISCO.   We believe that this work provides not only for a compelling fit to a disc model, but a deeper understanding of the underlying evolution. The results discussed here should be broadly applicable at late times to any TDE whose spectrum is dominated by a disc.

\section*{Acknowledgements}
We would like to thank our referee for an extremely detailed and helpful review, which greatly improved the presentation.   
This work is partially supported by the Hintze Family Charitable Trust and STFC grant ST/S000488/1.

\appendix
\section{Ray tracing algorithm}\label{appendixA}
\subsection*{Equations of motion}
 We use the Kerr metric in Boyer-Lindquist form, with co-ordinates $(t,\, r, \,\theta,  \,\phi)$. This metric has non-zero components $g_{00}, \, g_{rr},\, g_{\theta \theta},\, g_{\phi\phi},\, g_{0\phi} = g_{\phi0}$. Photon trajectories through the Kerr metric are characterised by two constants of motion, the angular momentum $L$, and energy $E$. These constants of motion are related to the photons 4-velocity $u^\mu$ by 
 \beq\label{L}
 L = p_\phi = g_{\phi\phi} u^{\phi} + g_{\phi 0} u^0 ,
 \eeq
 and
 \beq\label{E}
 E = -p_0 = - g_{00}u^0 - g_{0\phi}u^{\phi} .
 \eeq
 The photon 4-velocities $u^\mu$ are related to the co-ordinates $x^\mu$ and invariant line element $\text{d}\tau$ by
 \beq
 u^\mu = \frac{\text{d} x^\mu}{\text{d}\tau} .
 \eeq
 These two conservation laws can be re-written as equations of motion for the photon co-ordinates $t$ and $\phi$, explicitly: 
 \beq\label{teq}
 {{u^0}\over{E}} = \frac{\text{d}t}{\text{d}\tau'} = -\frac{lg_{\phi0} + g_{\phi\phi}}{g_{\phi\phi}g_{00} - g_{\phi0}^2},
 \eeq
and
 \beq\label{phieq}
 {{u^\phi}\over{E}} = \frac{\text{d}\phi}{\text{d}\tau'} = \frac{lg_{00} + g_{0\phi}}{g_{\phi\phi}g_{00} - g_{\phi0}^2},
 \eeq
 where $\tau' \equiv E\tau$ and $l \equiv L/E$. For the $r$ and $\theta$ co-ordinates we use the geodesic equations  
 \beq\label{req}
   \frac{\text{d}^2 r}{\text{d}\tau'^2} = - \Gamma^r_{\mu\nu}   \frac{\text{d}x^\mu}{\text{d}\tau'}   \frac{\text{d}x^\nu}{\text{d}\tau'} ,
 \eeq
 and
  \beq\label{thetaeq}
   \frac{\text{d}^2 \theta}{\text{d}\tau'^2} = - \Gamma^\theta_{\mu\nu}   \frac{\text{d}x^\mu}{\text{d}\tau'}   \frac{\text{d}x^\nu}{\text{d}\tau'} ,
 \eeq
where $\Gamma^{\mu}_{\nu \kappa} = \Gamma^{\mu}_{\kappa\nu}$ are the Christoffel coefficients for the Boyer-Lindquist Kerr metric. For an axi-symmetric metric the non-zero coefficients are $\Gamma^\star_{00}$, $\Gamma^\star_{rr}$, $\Gamma^\star_{\phi\phi}$, $\Gamma^\star_{\theta\theta}$, $\Gamma^\star_{\phi 0}$ and $\Gamma^\star_{\theta r}$, where $\star$ takes the place of $r$ and $\theta$ in equations (\ref{req}) and (\ref{thetaeq}) respectively. 

Photons propagating through the Kerr metric have an additional integral of motion, which arrises due to the vanishing norm of the photons 4-velocity  
\beq\label{norm}
g_{\mu\nu}  \frac{\text{d}x^\mu}{\text{d}\tau'}   \frac{\text{d}x^\nu}{\text{d}\tau'} = 0. 
\eeq

\subsection*{Photon initial condition}
We assume a distant observer orientated at an inclination angle $\theta_{\text{obs}}$ from the disc plane at a distance $D$. We set up an image plane perpendicular to the line of sight centred at $\phi = 0$ (Fig. \ref{fig1}), with image plane cartesian co-ordinates $(\alpha, \beta)$. A photon at an image plane co-ordinate  $(\alpha, \beta)$ has a corresponding spherical-polar co-ordinate $(r_i, \theta_i, \phi_i)$, given by  (Psaltis \& Johannsen 2012)
\begin{align}
r_i &= \left(D^2 + \alpha^2 + \beta^2\right)^{1/2}, \label{r0} \\
\cos\theta_i &=  r_i^{-1} \left( D \cos \theta_{\text{obs}} + \beta \sin\theta_{\text{obs}}\right) ,\\
\tan\phi_i &=   \alpha \left(D\sin\theta_{\text{obs}}- \beta \cos \theta_{\text{obs}} \right)^{-1}  .
\end{align}
The only photons which will contribute to the image have 3-momentum which is perpendicular to the image plane. This orthogonality condition uniquely specifies the initial photon 4-velocity (Psaltis \& Johannsen 2012)
\begin{align}
u^r_i &\equiv \left(\frac{\text{d} r}{\text{d}\tau'}\right)_{\text{obs}} = \frac{D}{r_i} , \\
u^\theta_i &\equiv \left(\frac{\text{d} \theta}{\text{d}\tau'}\right)_{\text{obs}}  = \frac{ D\left(D\cos \theta_{\text{obs}}+ \beta \sin \theta_{\text{obs}} \right) - r_i^2\cos\theta_{\text{obs}}  }{r_i^2 \left( r_i^2 - \left(D \cos\theta_{\text{obs}}   + \beta \sin\theta_{\text{obs}}  \right)^2\right)^{1/2}} , \\
u^\phi_i &\equiv \left(\frac{\text{d} \phi}{\text{d}\tau'}\right)_{\text{obs}} = \frac{- \alpha \sin\theta_{\text{obs}}}{\left( D \sin \theta_{\text{obs}} - \beta \cos \theta_{\text{obs}}\right)^2 + \alpha^2} . \label{up0}
\end{align}
We note that the normalisation of these 4-velocity components can all be scaled by an arbitrary factor $k$ without effecting the trajectories. 

The final component of the photons initial 4-velocity, $\left(\text{d} t /\text{d}\tau'\right)_\text{obs}$, can then be found from the initial 3-velocity and 3-position components (eq. \ref{r0} -- \ref{up0}) by the use of equation (\ref{norm}):
\beq\label{ut0}
 g_{00}(u^0_i)^2 + 2 g_{\phi 0}u^0_i u^\phi_i + g_{rr} (u^r_i)^2 + g_{\theta\theta} (u^\theta_i)^2 + g_{\phi\phi}(u^\phi_i)^2   
 = 0,
\eeq 
here the various metric components are evaluated at $(r_i, \theta_i, \phi_i)$.

\subsection*{Algorithm implementation }
The equations of motion (\ref{teq}, \ref{phieq}, \ref{req} \& \ref{thetaeq}) form the basis of our ray tracing algorithm. By writing 
\beq
  \frac{\text{d}^2 r}{\text{d}\tau'^2} =   \frac{\text{d} u^r}{\text{d}\tau'}, 
\eeq
and 
\beq
  \frac{\text{d}^2 \theta}{\text{d}\tau'^2} =   \frac{\text{d} u^\theta}{\text{d}\tau'},
\eeq
equations (\ref{teq}--\ref{thetaeq}) can be expressed as four coupled first order differential equations for the variables $(t, \,\phi,\, u^r,\, u^\theta)$. These four equations, together with the definitions
\beq\label{ureq}
\frac{\text{d} r}{\text{d}\tau'} =  u^r ,
 \eeq
 and
 \beq\label{uteq}
 \frac{\text{d} \theta}{\text{d}\tau'} =  u^\theta,
 \eeq
 completely specify the photons trajectory. We solve these  six (\ref{teq} -- \ref{thetaeq}, \ref{ureq}, \ref{uteq})  coupled first order differential equations using a fourth order Runge-Kutta integrator. The initial condition is that  of a photon at the image plane with position and 4-velocity given by equations (\ref{r0} -- \ref{ut0}), this photon is then propagated backwards until it crosses the disc plane, at some radius $r_f$. 
 
 We employ a variable time step $\delta \tau'$, set as a fixed fraction $h$ of the fastest changing variable
 \beq
 \delta \tau' = h \times \text{min} \left[ r  \left(\frac{\text{d}r}{\text{d}\tau'}\right)^{-1}, \, \theta  \left(\frac{\text{d}\theta}{\text{d}\tau'}\right)^{-1},\, \phi  \left(\frac{\text{d}\phi}{\text{d}\tau'}\right)^{-1} \right]  .
 \eeq

To determine the appropriate size of the fixed step size $h$ we require a measure of the accuracy of the algorithm. The final integral of motion (equation \ref{norm}) is useful for this purpose. Errors propagating throughout the photons trajectory will cause the norm of the photons 4-velocity to deviate from zero. We therefore define the parameter $\Delta$ by 
 \begin{multline}
 \Delta = 1 + \Bigg[g_{rr} \left(\frac{\text{d}r}{\text{d}\tau'}\right)^2 + g_{\theta\theta} \left(\frac{\text{d}\theta}{\text{d}\tau'}\right)^2 + g_{\phi\phi} \left(\frac{\text{d}\phi}{\text{d}\tau'}\right)^2   \\ 
 + 2 g_{\phi 0} \left(\frac{\text{d}t}{\text{d}\tau'}\right)\left(\frac{\text{d}\phi}{\text{d}\tau'}\right)\Bigg] \Bigg/ {g_{00} \left(\frac{\text{d}t}{\text{d}\tau'}\right)^2} ,
 \end{multline}
which would satisfy  $\Delta = 0$ for an error free integration. We set the fixed step size $h$ by requiring that 
\beq
 \Delta < 10^{-7},
 \eeq
for all photon trajectories. This was found empirically to be satisfied by 
\beq
h = 5 \times 10^{-3}. 
\eeq

\subsection*{Photon red shift and the observed spectrum}
The observed disc spectrum $F_\nu$ is given by the double integral 
\beq\label{Aflux}
F_\nu = {{1}\over{D^2}}  \iint_{\cal S} {f^3 B_\nu (\nu_o/f , T)}\,  \text{d}\alpha \, \text{d}\beta ,
\eeq
where the disc temperature $T(r,t)$ is given by equation (\ref{temperature}), and
the photon red shift factor $f$ is given by 
\beq
f = \frac{\nu_o}{\nu_e} = \frac{1}{U^{0}} \left[ 1+ \frac{p_{\phi}}{p_0} \Omega \right]^{-1} .
\eeq
Here $U^0$ and $\Omega = U^\phi / U^0$ are 4-velocity components of the rotating  disc fluid, and $p_\phi$ and $p_0$ are photon 4-momentum components.  The ratio $p_\phi / p_0$ is a constant of motion for a photon propagating through the Kerr metric, and is equal to $-l$ in the notation of our algorithm. As a conserved quantity, $l$ can  be calculated from the initial conditions and equations (\ref{L} \& \ref{E}). The quantities $U^0$ and $\Omega$ are functions of the black hole spin $a$, and the radius at which the photon was emitted. This radius corresponds to the end point of the ray-tracing algorithm  $r_f$.  

Starting from a finely spaced grid of points $(\alpha, \beta)$ in the image plane, we trace the geodesics of each photon back to the disc plane, recording $(r_f, l)$ for each photon.  The parameter $r_f$ allows the disc temperature $T$ to be calculated at a given time $t$ (equation \ref{temperature}). The parameters $r_f$ and $l$ together uniquely define the red-shift factor $f$. The integrand $f^3B_\nu (\nu_o/f , T)$  can therefore be calculated at every grid point in the image plane. The integral (\ref{Aflux}) was then calculated using a 2-dimensional Simpsons algorithm. 

\section{Fitting procedure and parameter degeneracies}\label{appendixB}
\begin{figure}
  \includegraphics[width=.45\textwidth]{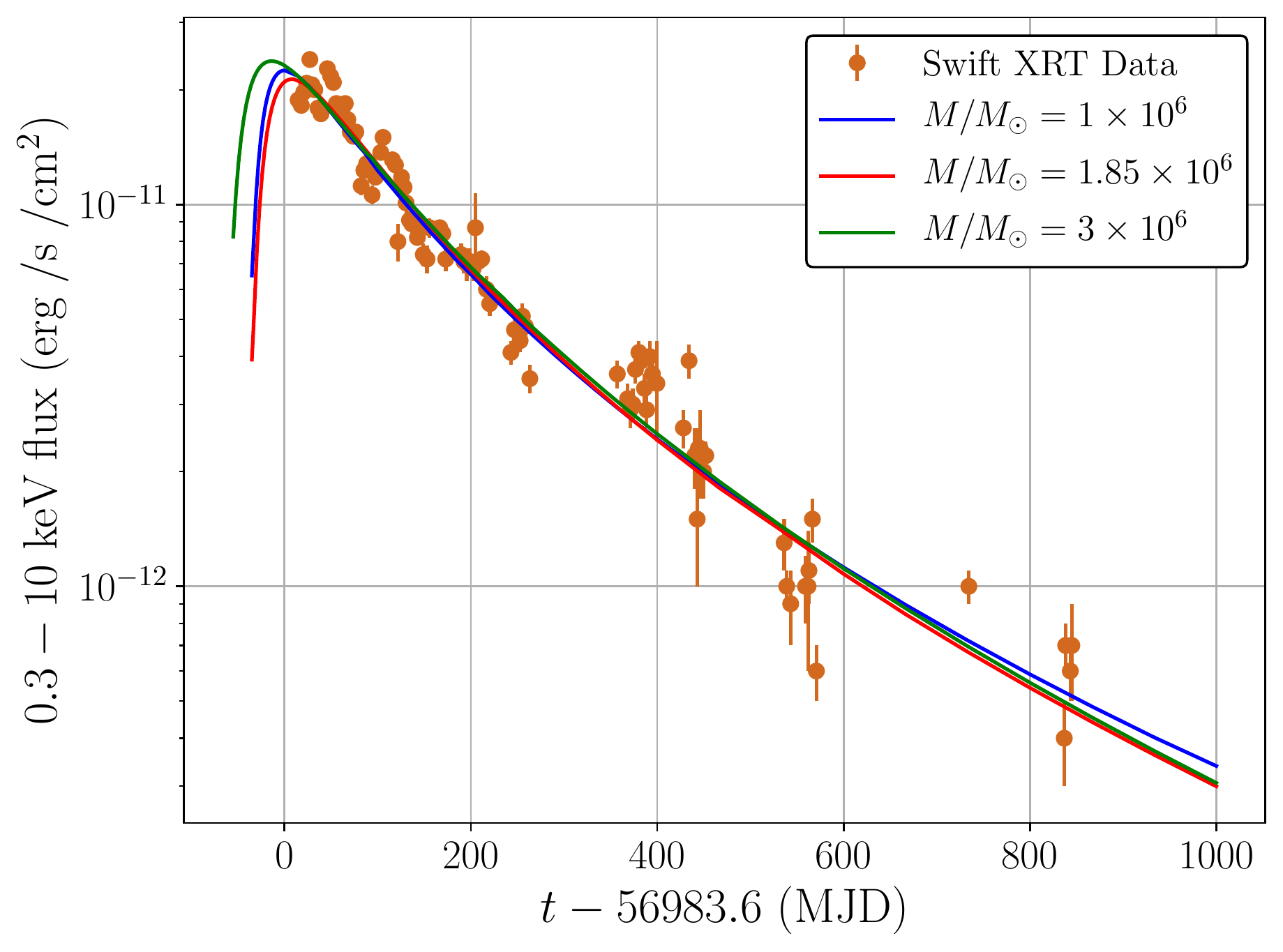} 
 \caption{Best-fit X-ray light curves for different black hole masses, ignoring the late-time \Asa\ UV data. The difficulty in tightly constraining system parameters exclusively with the X-ray data is evident. Black hole masses for each curve are shown.  }
 \label{xraydegeneracy}
\end{figure}
\begin{figure}
  \includegraphics[width=.45\textwidth]{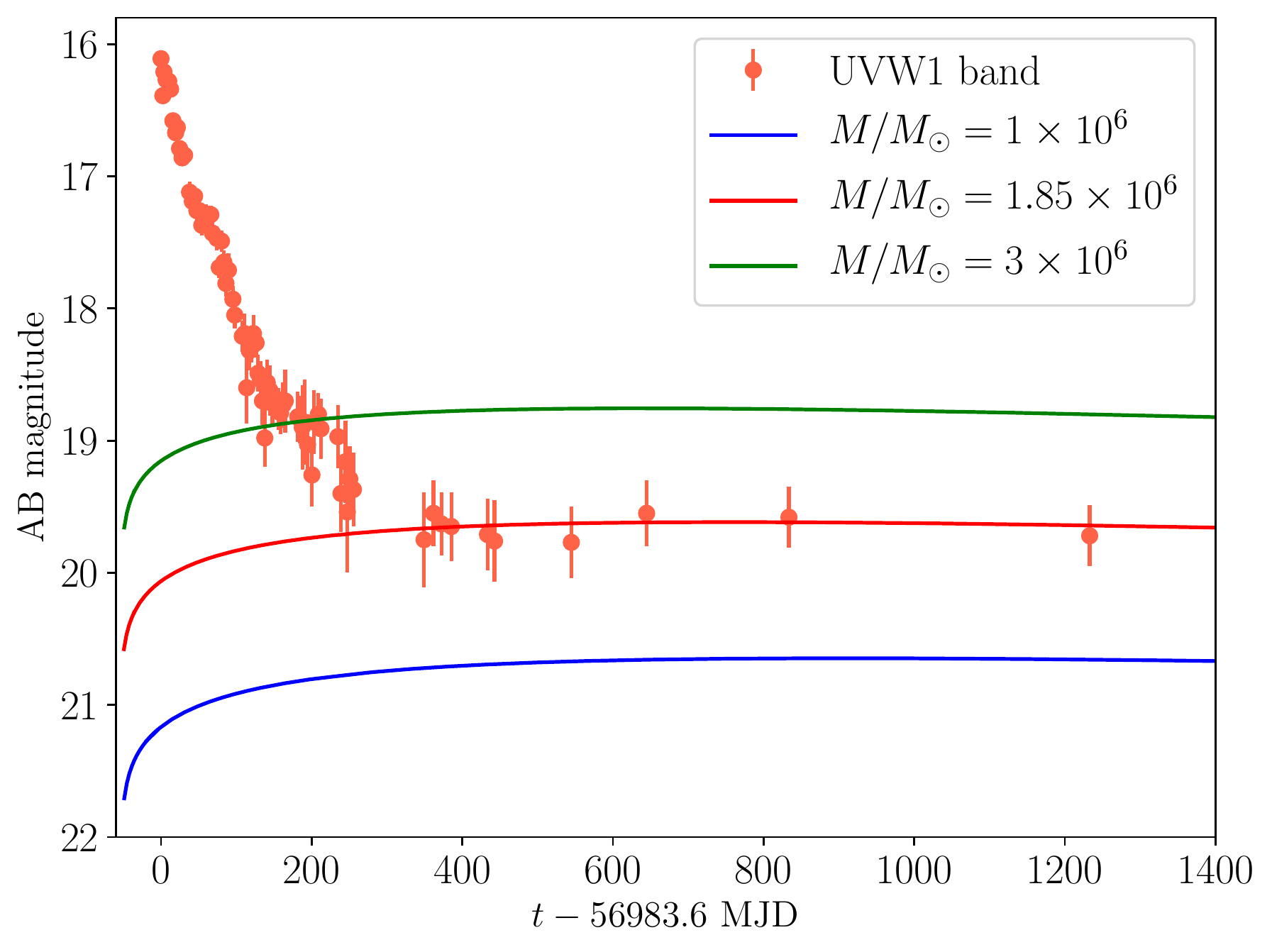} 
 \caption{The disc light curves in the UVW1 band produced by the degenerate X-ray light curve parameters of Fig. \ref{xraydegeneracy}. Although the X-ray light curves are completely degenerate, the corresponding UVW1 fluxes significantly differ from the observed values except for narrow ranges of the black hole mass.   }
 \label{xraydegeneracyFUV}
\end{figure}
\begin{figure}
  \includegraphics[width=.45\textwidth]{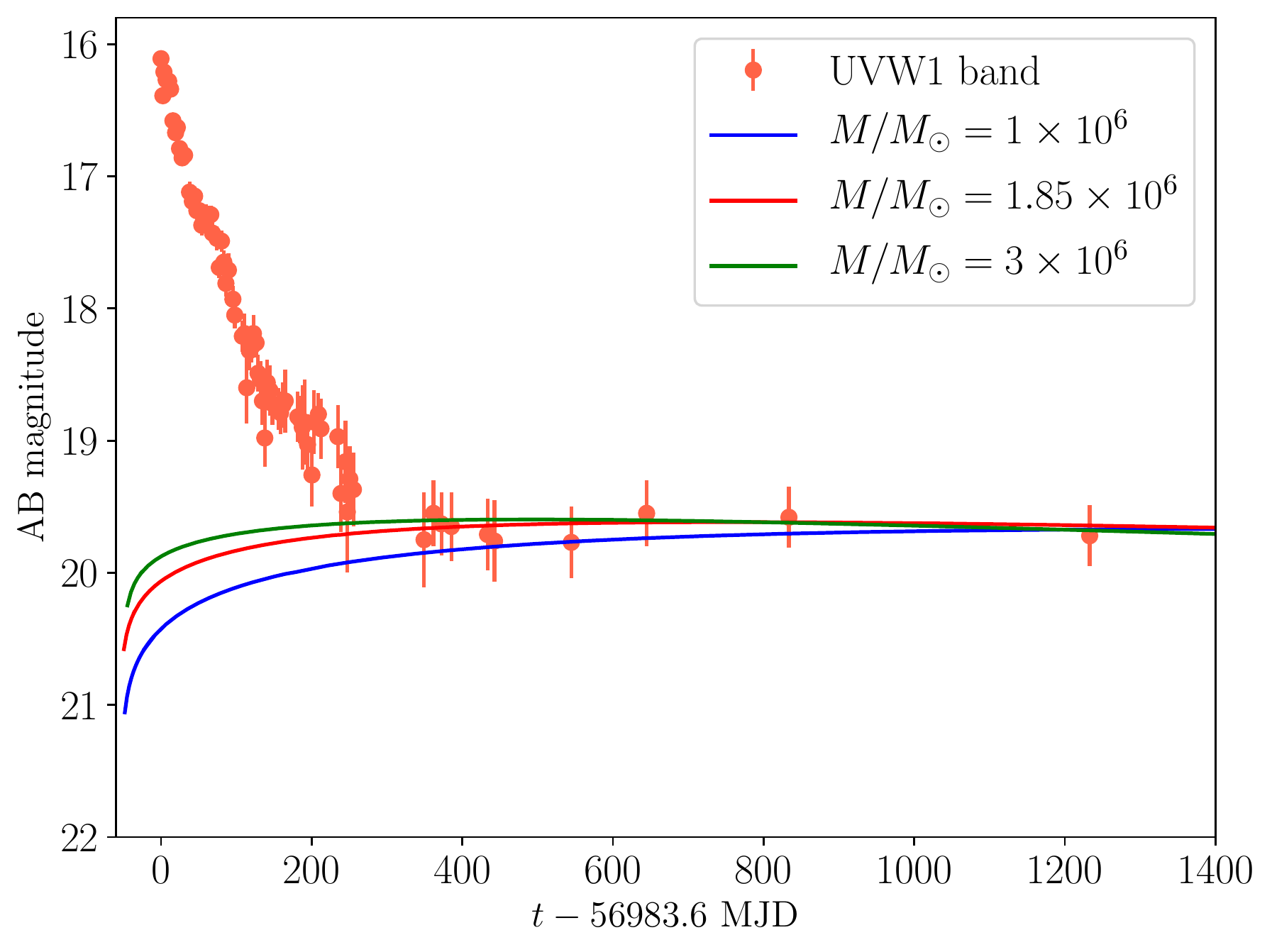} 
 \caption{Best-fit UVW1 light curves for different black hole masses, ignoring the \Asa\ X-ray data.  The black hole masses for each curve are denoted on the plot. Like the X-ray light curves, the late-time UVW1 light curves of \Asa\ can individually be fit by a wide range of system parameters. }
 \label{FUVdegeneracy}
\end{figure}
\begin{figure}
  \includegraphics[width=.45\textwidth]{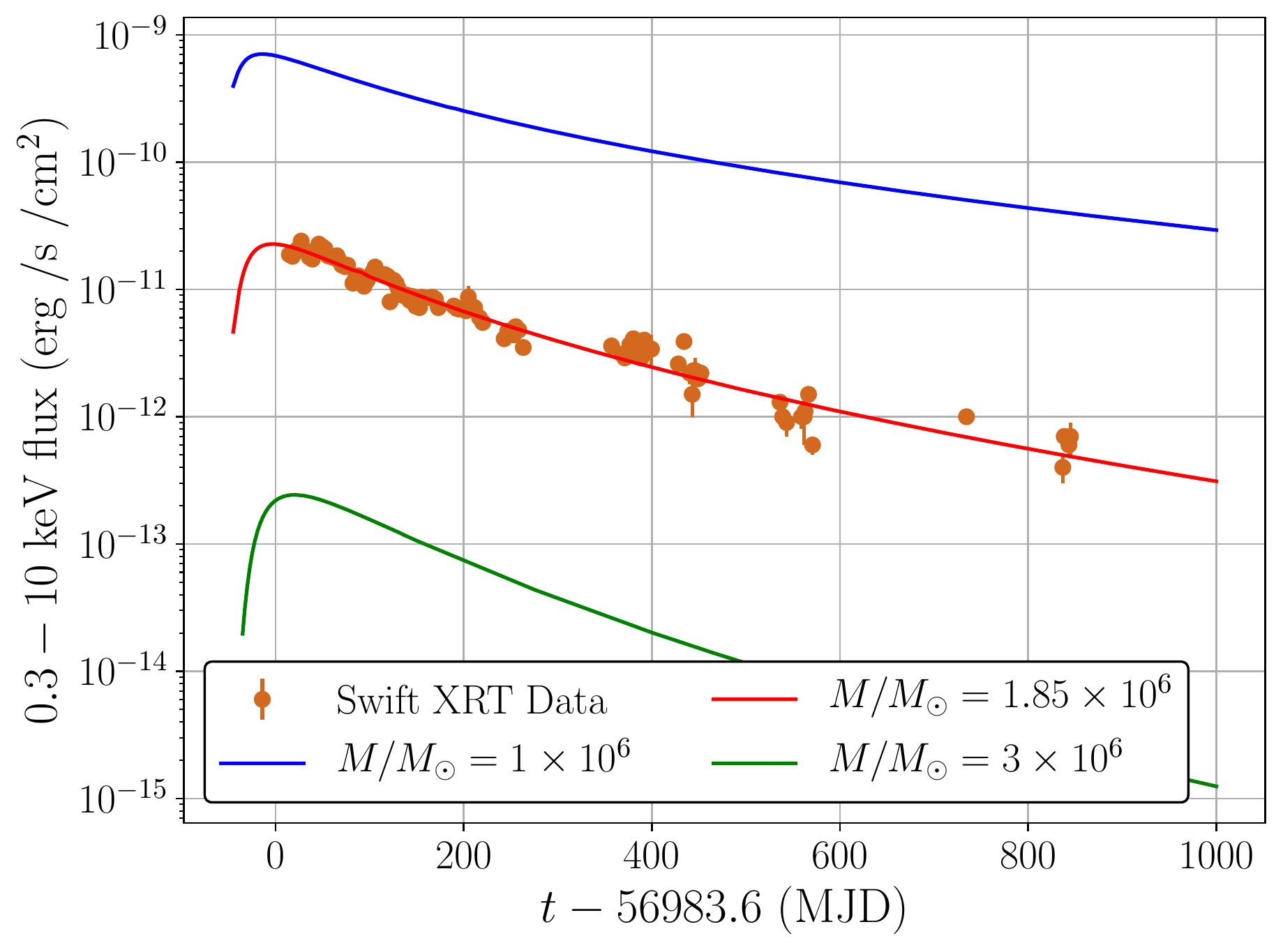} 
 \caption{The X-ray light curves produced by the degenerate UVW1 light curve parameters of Fig. \ref{FUVdegeneracy}. Degenerate UV light curves here produce orders of magnitude differences in observed X-ray fluxes.   }
 \label{FUVdegeneracyXRAY}
\end{figure}
\begin{figure}
  \includegraphics[width=.45\textwidth]{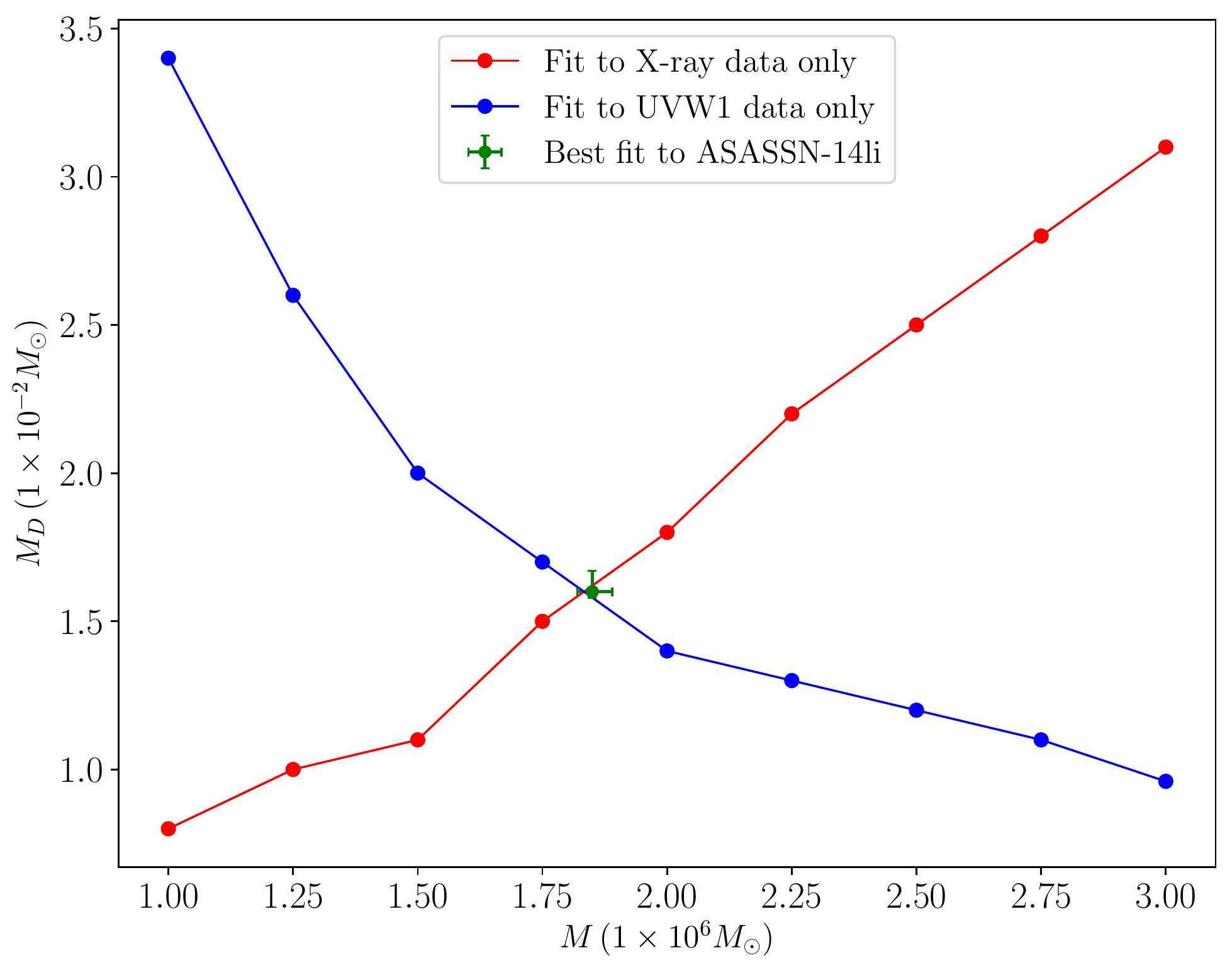} 
 \caption{The best fit disc mass as a function of black hole mass for individual light curve fits to the UVW1 band (blue curve) and X-ray band (red curve). The parameter dependencies have opposite trends for the different light curves, meaning that we can tightly constrain the system parameters by using both sets of data. The errors bars on the \Asa\ parameters are representative of the error bars on all the points.       }
 \label{contours}
\end{figure}

{We fit four free parameters to the observed \Asa\ light curves: the time offset $t_D$, the viscous timescale $t_v$, the disc mass $M_d$, and the black hole mass $M$. The time offset and viscous timescale of the evolving disc are constrained by minimising the reduced chi-squared of the \Asa\ X-ray light curve, which is strongly time varying.  Unlike the temporal parameters, the black hole and disc masses cannot be constrained by either the X-ray or UV observations separately.  This is shown in Figs.\ \ref{xraydegeneracy} and \ref{FUVdegeneracy}.  

The black hole and disc masses may be constrained, however, by simultaneously fitting the UV and X-ray light curves.  Figures \ref{xraydegeneracyFUV} and \ref{FUVdegeneracyXRAY} demonstrate that disc parameters which equally well fit the UV band light curve, produce X-ray light curves which differ from the observed fluxes by orders of magnitude, and vice-versa.   Fitting separately to the UV and X-ray bands produce two different curves between disc and black hole mass, their unique intersection allows both to the determined (Fig. \ref{contours}).     To fit the black hole and disc masses we minimised the reduced chi-squared of the X-ray light curve, subject to the constraint that the late time UV flux plateau must have the correct magnitude.   Note that the X-ray light curve is a more extensive data set with more complex temporal structure, and is therefore more constraining.  

Fluctuations in the data mean that our best fit X-ray light curve still has a large reduced chi-squared, which makes estimating the error ranges on our best fit parameters problematic.   Once a best fit had been found, our approach has been to formally enlarge the quoted error bars on the X-ray data  (thereby accounting for the fluctuations in a statistical sense) by a constant factor until the best fit reduced chi-squared is equal to unity.   Using this manipulated data, the quoted error ranges are found by computing the $\Delta \chi^2 = 3$ contours of the X-ray light curves, while ensuring that the late time UV flux was correctly reproduced.   These should be considered very conservative error ranges.  }

\section{Edge on viewer orientation}\label{appendixC}
 We require the asymptotic expansions of the integrals $I$ (flux integral) \& $J$ (luminosity integral), where 
 \beq
I  = \iint_{\cal S}  \exp\left(- {h\nu_o}/{k_B \widetilde T} \right) ~ {\text{d}\alpha \, \text{d} \beta} ,
\eeq
and
\beq
J = \int_{\nu_l}^\infty I\, \text{d}\nu_o .
\eeq
The effective temperature is defined here, as in the text (eq. \ref{Teff}),
\beq
\widetilde T(\alpha, \beta, t) = f(\alpha, \beta)\,T(r(\alpha,\beta), t) .
\eeq
In particular, we are interested in the properties of the solutions in the limit 
\beq
h \nu_l \gg k_B \widetilde T ,
\eeq
when the disc is observed at an angle $\theta_{\rm obs}$ not close to face-on. When a relativistic disc is observed at angles close to edge-on, the large rotational velocities in the innermost disc regions  result in strong doppler blue shifting of the observed photons emergent from one half of the disc, and similarly strong doppler red shifting of the remaining photons.    Therefore, the frequency ratio function $f(\alpha, \beta)$ will peak strongly about a `hot spot' surrounding the most strongly blue shifted material. We must therefore modify the asymptotic expansion techniques of sections \ref{QWlim} and \ref{VSexpansion} to take account of this different observing geometry. 

We define the relevant function 
\beq
K(\alpha, \beta ,t) \equiv {{h\nu_o}\over{k_B \widetilde T}} ,
\eeq
and proceed to expand $K$ about  the co-ordinates of the discs hotspot $(\alpha, \beta) = (X_0, Y_0)$, where 
\beq  
\widetilde T_p \equiv \widetilde T(X_0, Y_0) = {\rm max}\left[ \widetilde T(\alpha, \beta) \right] .
\eeq
The exact form of the expansion will then depend on the nature of the ISCO stress. Crucially for our purposes, the leading order term in the expansion will have time dependence as in \S \ref{QWlim}
\beq
K_0 \equiv {h\nu_l \over k_B \widetilde T_p }  \propto t^{n/4} .
\eeq
Finally, the effects of  gravitational lensing of the observed   inner disc geometry will be accounted for with 
\beq
{\rm d}\alpha \, {\rm d}\beta = F(X,Y)\, {\rm d}X{\rm d}Y,
\eeq
here $F$ is an unspecified function of coordinates (but, importantly, not $\nu$) and ${\rm d}X{\rm d}Y$ the local area element (Balbus 2014). 
\subsection*{Finite ISCO stress -- hot spot on an inner boundary}
If the stress is finite at the ISCO, then the disc temperature profile will peak at its innermost edge. The leading order expansion of $K$ is then given by
\beq
K =  K_0 +  K_X (X-X_0) + {{1}\over{2}} K_{YY} (Y-Y_0)^2 + ...
\eeq
where $X$ and $Y$ have been appropriately chosen so that any cross-terms vanish. A subscript  $X$ or $Y$ here denotes a partial derivate with respect to the relevant variable, with all other variables held constant. The spectrum integral is then 
 \begin{multline}
I  = \exp\left(- K_0 \right) \int_{-\infty}^{\infty} \,\int_{X_0}^{\infty} F(X,Y)  \exp\left[ - K_X (X-X_0) \right] \\ 
\exp\left[- {{1}\over{2}} K_{YY} (Y-Y_0)^2\right]   ~ {\text{d}X \, \text{d} Y} ,
\end{multline}
which has a leading order solution
\begin{multline}
I = F(X_0, Y_0) \frac{X_0Y_0}{\xi_X h\nu_o } \sqrt{\frac{2\pi}{\xi_{YY} h\nu_o }}\\ \, (k_B \widetilde T_p)^{3/2}  \exp\left(- {h\nu_o}/k_B \widetilde T_p \right) .
\end{multline}
Here we have defined magnitude constants 
\beq
\xi_X \equiv X_0 K_X, \quad \xi_{YY} \equiv Y_0^2 K_{YY}. 
\eeq
The leading order luminosity integral then simply follows
\begin{align}
J &\propto (k_B \widetilde T_p)^{-5/2}  \exp\left(- {h\nu_l}/k_B \widetilde T_p \right) , \\
J &\propto t^{-5n/8} \exp\left(-(t/t_X)^{n/4} \right) ,
\end{align}
with, for a finite ISCO stress disc, $n \simeq 0.8$. The only difference resulting from changing the disc orientation from face-on to edge-on is  a slight modification to the leading power-law exponent, from $4n/8$ (face-on) to $5n/8$ (edge-on).  

 \subsection*{Vanishing ISCO stress -- exterior hot spot}
 As in \S\ref{VSexpansion}, when the disc stress vanishes at the ISCO, the disc temperature maximum occurs at a disc radius {exterior} to the ISCO. The relevant expansion is then  (employing  the same subscript notation as above)
\beq
K =  K_0 + {{1}\over{2}}  K_{XX} (X-X_0)^2 + {{1}\over{2}}  K_{YY} (Y-Y_0)^2 + ...
\eeq
This results in a leading order flux integral
 \begin{multline}
I  = \exp\left(- K_0 \right) \int_{-\infty}^{\infty} \,\int_{-\infty}^{\infty} F(X,Y)\, \times \\ \exp\left[ - {{1}\over{2}} \left( K_{XX} (X-X_0)^2 + K_{YY} (Y-Y_0)^2\right)\right]   ~ {\text{d}X \, \text{d} Y} ,
\end{multline}
which is solved to give
\begin{multline}
I = F(X_0, Y_0) X_0Y_0 \sqrt{\frac{2\pi}{\xi_{XX} h\nu_o }} \sqrt{\frac{2\pi}{\xi_{YY} h\nu_o }} \, \times \\ (k_B \widetilde T_p)^{-1}  \exp\left(- {h\nu_o}/ k_B \widetilde T_p \right) ,
\end{multline}
where
\beq
\xi_{XX} \equiv X_0^2 K_{XX}, \quad \xi_{YY} \equiv Y_0^2 K_{YY}. 
\eeq
The leading order luminosity integral then follows simply:
\begin{align}
J &\propto (k_B \widetilde T_p)^{-2}  \exp\left(- {h\nu_l}/ k_B \widetilde T_p\right), \\
J &\propto t^{-n/2} \exp\left(-(t/t_X)^{n/4} \right),
\end{align}
with $n \simeq 1.2$ for a vanishing ISCO stress disc. As for a finite ISCO stress disc, modifying the disc-observer orientation angle only results in a slightly changed leading power-law exponent.

\label{lastpage}

\end{document}